\documentclass[%
reprint,
superscriptaddress,
nofootinbib,
 amsmath,amssymb,
 aps,
]{revtex4-1}

\usepackage{graphicx}% Include figure files
\usepackage{dcolumn}% Align table columns on decimal point
\usepackage{bm}% bold math
\usepackage[utf8]{inputenc}
\usepackage[colorlinks]{hyperref}% add hypertext capabilities
\usepackage{physics}
\usepackage{qcircuit}
\usepackage{tikz}
\usetikzlibrary{matrix,chains,positioning,decorations.pathreplacing,arrows}
\usepackage{mathtools}
\usepackage{subfig}
\usepackage{float}
\usepackage{pgfplots}
\usepackage{xcolor}
\pgfplotsset{compat=newest}
\usepgfplotslibrary{groupplots}
\usepgfplotslibrary{dateplot}

\definecolor{g85}{gray}{0.333}
\definecolor{g170}{gray}{0.666}

\definecolor{g102}{gray}{0.4}
\definecolor{g16}{gray}{0.0636}

\definecolor{g204}{gray}{0.8}
\definecolor{g28}{gray}{0.11}
\definecolor{g40}{gray}{0.16}

\definecolor{g32}{gray}{0.127}
\definecolor{g162}{gray}{0.6366}
\definecolor{g65}{gray}{0.25}
\definecolor{g211}{gray}{0.827}

\definecolor{g242}{gray}{0.949}
\definecolor{g60}{gray}{0.235}
\definecolor{g105}{gray}{0.411}

\definecolor{g17}{gray}{0.066}
\definecolor{g230}{gray}{0.901}
\definecolor{g196}{gray}{0.768}
\definecolor{g110}{gray}{0.431}

\begin{document}

%\title{Quantum neuron with continuous phases}% Force line breaks with \\
\title{Quantum computing model of an artificial neuron with continuously valued input data}

\author{Stefano Mangini}
\email{stefano.mangini01@universitadipavia.it}
\affiliation{Dipartimento di Fisica, Università di Pavia, via Bassi 6, I-27100, Pavia, Italy}
\author{Francesco Tacchino}
 \altaffiliation[Present address ]{
 \textit{IBM Research GmbH, Zurich Research Laboratory, S\"{a}umerstrasse 4, CH-8803 R\"{u}schlikon, Switzerland}}%Lines break automatically or can be forced with \\
\affiliation{Dipartimento di Fisica, Università di Pavia, via Bassi 6, I-27100, Pavia, Italy}

\author{Dario Gerace}
\affiliation{Dipartimento di Fisica, Università di Pavia, via Bassi 6, I-27100, Pavia, Italy}

\author{Chiara Macchiavello}
\affiliation{Dipartimento di Fisica, Università di Pavia, via Bassi 6, I-27100, Pavia, Italy}
\affiliation{INFN Sezione di Pavia, via Bassi 6, I-27100, Pavia, Italy}
\affiliation{CNR-INO, largo E. Fermi 6, I-50125, Firenze, Italy}

\author{Daniele Bajoni}
\affiliation{Dipartimento di Ingegneria Industriale e dell’Informazione, Università di Pavia, via Ferrata 1, I-27100, Pavia, Italy}

\date{\today}% It is always \today, today,
             %  but any date may be explicitly specified

\begin{abstract}
{ Artificial neural networks have been proposed as potential algorithms that could benefit from being implemented and run on quantum computers. In particular, they hold promise to greatly enhance Artificial Intelligence tasks, such as image elaboration or pattern recognition. The elementary building block of a neural network is an artificial neuron, i.e. a computational unit performing simple mathematical operations on a set of data in the form of an input vector. Here we show how the design for the implementation of a previously introduced quantum artificial neuron [npj Quant. Inf. {\bf 5}, 26], which fully exploits the use of superposition states to encode binary valued input data, can be further generalized to accept continuous- instead of discrete-valued input vectors, without increasing the number of qubits. This further step is crucial to allow for a direct application of an automatic differentiation learning procedure, which would not be compatible with binary-valued data encoding. } 
\end{abstract}

% \keywords{Suggested keywords}%Use showkeys class option if keyword
                              %display desired
\maketitle

%\tableofcontents

\section{Introduction}
\label{sec:Intro}

{Quantum computers hold the promise to greatly enhance the computational power of not-so-distant in future computing machines \cite{QuantumSupremacy, NISQ}. In particular, improving machine learning techniques by means of quantum computers is the essence of the raising field of the field of Quantum Machine Learning \cite{biamonte_quantum_2017, Dunjko_Briegel_2018, benedetti_parameterized_2019}. Several models for the quantum computing version of artificial neurons have been proposed \cite{schuld_simulating_2015, wiebe_quantum_2016, cao_quantum_2017, Tacchino0, torrontegui_unitary_2019,  kristensen_artificial_2019}, together with novel quantum machine learning techniques implementing classification tasks \cite{havlicek_supervised_2019, schuld_implementing_2017, schuld_quantum_2019}, quantum autoencoders \cite{Romero_2017, Lamata_2018}, and quantum convolutional networks \cite{henderson_quanvolutional_2020, cong_quantum_2019}, to give a non-exhaustive list.}

In this context, quantum signal processing leverages the capabilities of quantum computers to represent and elaborate exponentially large arrays of numbers, and it could be used for enhanced pattern recognition tasks, i.e. going beyond the capabilities of classical computing machines \cite{Cohen_Quantum_Image_Processing_IEEE_2003}. In these regards the development of Neural Networks dedicated for quantum computers \cite{Schuld_Petruccione_review_2014} is of fundamental importance, due to the preponderance of this type of classical algorithms in image processing \cite{Zurada:intro_ANN_1992,Rojas_ANN_Introduction}. 

In the commonly accepted terminology of graph theory, neural networks are directed acyclic graphs (DAG), i.e., a collection of nodes where information flows only in one direction, without any loop. Each node is generally defined an artificial neuron, i.e., a simplified mathematical model of natural neurons. In practice, it consists of an object function that takes some input data, processes them using some internal parameters (defined weights), and eventually gives an output value. In their simplest form, the so called McCulloch-Pitts neurons \cite{McCulloch_Pitts_1943} only deal with binary values, while in the most common and most useful form, named perceptron \cite{Rosenblatt}, they accept real, continuously valued inputs and weights.

Continuous inputs are not possible in conventional, digital computers, and these are usually rendered by using bit strings: a grey scale image pixel is for instance usually rendered in natural numbers on a scale from 0 to 255 using 8-bit binary strings. Some approaches propose to use a similar representation in quantum computers by assigning several qubits per value \cite{Li2013,Hirota_quantum_images_polynomial_preparation,Latorre_image_compression_entanglement}.
However, these approaches are particularly wasteful, especially in light of the fact that quantum mechanical wavefunctions can be inherently represented as continuously valued vectors.

 A previous work \cite{Tacchino0} introduced a model for a quantum circuit mimicking a McCulloch-Pitts neuron. Here we generalize that model to the case of a quantum circuit accepting also continuously valued input vectors. We thus present a model for a continuous quantum neuron which, as we will see, can be used for pattern recognition in greyscale images without the need to increase the number of  qubits to be employed. This  represents a further memory advantage with respect to classical computation, where an increase in the number of encoding bits is required to deal with continuous numbers. We employ a phase-based encoding, and show that it is particularly resilient to noise.
 
 Differently from classical perceptron models, artificial quantum neurons as described, e.g., in Ref.~\cite{Tacchino0} can be used to classify linearly non separable sets. In the continuously valued case, we thus harness the behaviour of our quantum perceptron model to show its ability to correctly classify several notable cases of linearly non separable sets. Furthermore, we test this quantum artificial neuron for digit recognition on the MNIST dataset \cite{ClassificationMNIST}, with remarkably good results.
 We further stress that the present generalization of the binary-valued artificial neuron model is a crucial step in view of fully exploiting the great potential allowed from automatic differentiation such as gradient descent. These techniques are commonly employed, e.g., in supervised and unsupervised learning procedures, and would be impossible to be applied to the oversimplified McCulloch-Pitts neuron model.

\section{Continuously valued quantum neuron model}
\label{sec:CQN}

\subsection{The algorithm}
\label{subsec:Algo}

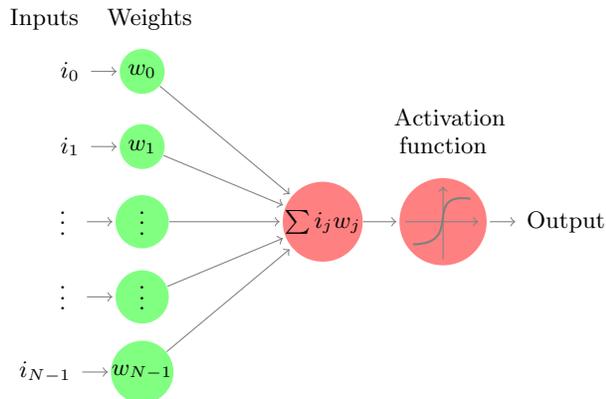
\begin{figure}
\def\layersep{2.4cm}

\begin{tikzpicture}[shorten >=1pt,draw=black!50, node distance=\layersep]
    \tikzstyle{every pin edge}=[<-,shorten <=1pt]
    \tikzstyle{neuron}=[circle,fill=black!25,minimum size=17pt,inner sep=0pt]
    \tikzstyle{input neuron}=[neuron, fill=green!50];
    \tikzstyle{output neuron}=[neuron, fill=red!50];
    \tikzstyle{annot} = [text width=4em, text centered]
    
     \node[input neuron, pin=left:$i_0$] (i0) at (0, 0) {$w_0$};
     \node[input neuron, pin=left:$i_1$] (i1) at (0,-1) {$w_1$};
     \node[input neuron, pin=left:\raisebox{0.55 em}{$\vdots_{\ }$}](i2) at (0,-2) {\raisebox{0.55 em}{$\vdots$}};
     \node[input neuron, pin=left:\raisebox{0.55 em}{$\vdots_{\ }$}] (i3) at (0,-3) {\raisebox{0.55 em}{$\vdots$}};
     \node[input neuron, pin=left:$i_{N-1}$] (i4) at (0,-4) {$w_{N-1}$};
     
      %Draw the output layer node
    \node[output neuron] (O) at (\layersep, -2) {$\sum i_j w_j $};
    
     %Draw the output layer node
    \node[output neuron,pin={[pin edge={->}]right:Output}, minimum size=33pt] (O2) at (4, -2) {};
     \draw[thick] (3.6,-2.3) .. controls (4.3,-2.3) and (3.7,-1.6) .. (4.4,-1.7);
     \draw[->] (3.5,-2) --(4.5,-2);
     \draw[->] (4,-2.5) --(4,-1.5);
     
      %Connect every node in the hidden layer with the output layer
	\path[->] (O) edge (O2);     
     \path[->] (i0) edge (O);
     \path[->] (i1) edge (O);
     \path[->] (i2) edge (O);
     \path[->] (i3) edge (O);
     \path[->] (i4) edge (O);
     
     % Annotate the layers
	 \node[annot,above of= O2, node distance=1.2cm] {Activation function};
	 \node[annot] at (-1.3, 0.7) {\small Inputs};
	 \node[annot] at (0.1, 0.7) {\small Weights};
\end{tikzpicture}
\caption{{\bf Scheme of a classical perceptron model.} The artificial neuron evaluates a weighted sum between the input vector, $\vec{i}$, and the weight vector, $\vec{w}$, followed by an activation function, which determines the actual output of the neuron.}
\label{fig:classical_perceptron}
\end{figure}

{Let us consider a perceptron model with real valued input and weight vectors}, which are respectively indicated as $\vec{i}$ and $\vec{w}$, such that $i_j, w_j \in \mathbb{R}$. A schematic representation of the classical perceptron is reported in Fig.~\ref{fig:classical_perceptron}.

% End of code

Similarly, we define a model of a quantum neuron capable of accepting continuously valued input and weight vectors, by extending a previous proposal for the quantum computing model of an artificial neuron only accepting binary valued input data \cite{Tacchino0}. In order to encode data on a quantum state, we make use of a phase encoding. Given an input $\bm{\theta}=(\theta_0, \hdots, \theta_{N-1})$ with $\theta_i \in [0, \pi]$, which consists of the classical data to be analyzed, we consider the vector:
\begin{equation}
\vec{i}=(e^{i \theta_0}, e^{i \theta_2}, \cdots, e^{i\theta_{N-1}})\ , 
\label{eq:quantum_input}
\end{equation}
which we will be referring to as the input vector in the following. With this input vector we define the input quantum state of $n=\log_2 N$ qubits:
\begin{equation}
\ket{\psi_i} = \frac{1}{2^{n/2}}\sum_{k=0}^{2^{n}-1} i_k \ket{k}\ ,
\label{eq:input}
\end{equation}
where the states $\ket{k}$ denote the computational basis states of $n$ qubits ordered by increasing binary representation, $\{\ket{00\hdots 0}, \ket{00\hdots 1},\cdots,\ket{11\hdots 1}\}$.
Since we are dealing with an artificial neuron, we have to properly encode another vector, which represents the weights in the form $\bm{\phi}=(\phi_0, \hdots, \phi_{N-1})$ with $\phi_i \in [0,\pi]$, i.e. the corresponding vector:
\begin{equation}
\vec{w}=(e^{i \phi_0}, e^{i \phi_2}, \cdots, e^{i \phi_{N-1}})
\label{eq:quantum_weight}
\end{equation}
which in turn defines the  weight quantum state:
\begin{equation}
\ket{\psi_w} = \frac{1}{2^{n/2}}\sum_{k=0}^{2^{n}-1} w_k \ket{k}\ .
\label{eq:weight}
\end{equation}

Notice that \eqref{eq:input} and \eqref{eq:weight} have the same structure, i.e. they consist of an equally weighted superposition of all  the computational basis states, although with varying phases. By means of such encoding scheme, we can fully exploit the exponentially large dimension of the $n$ qubits Hilbert space, i.e.,  by only using $n$ qubits it is evidently possible to encode and analyze data of dimension $N=2^n$. Due to global phase invariance, the number of actual independent phases is $2^n-1$, which does not spoil the overall efficiency of the algorithm, as it will be shown. We also notice that the class of states represented as $\frac{1}{2^{n/2}}\sum_i e^{i\alpha_i}\ket{i}$, as \eqref{eq:input} and \eqref{eq:weight} are known as locally maximally entanglable (LME) states, as introduced in Ref.~\cite{Kraus}.

Having defined the input and weight quantum states, their similarity is estimated by considering the inner product
\begin{eqnarray}
\braket{\psi_w}{\psi_i} & := & \frac{1}{2^n}\ \sum_{k,j=0}^{2^n-1} i_k w^*_j \braket{j}{k}  \nonumber \\
& = & \frac{1}{2^n}\ \vec{i} \cdot \vec{w^*}  \\
& = & \frac{1}{2^n}\left(e^{i(\theta_0 - \phi_0)} + \cdots + e^{i(\theta_{2^n-1} - \phi_{2^n-1})}\right)\ ,  \nonumber
\label{eq:innerprod}
\end{eqnarray}
which corresponds to evaluating the scalar product between the input vector in Eq.~\eqref{eq:quantum_input} and the conjugated of the weight vector in Eq.~\eqref{eq:quantum_weight}, $\vec{w}^*$, similarly to the classical perceptron algorithm. Since probabilities in quantum mechanics are represented by the squared modulus of wavefunction amplitudes, we consider $|\braket{\psi_w}{\psi_i}|^2$, which is explicitly given as (see App.~\ref{app:A}):
\begin{equation}
\label{eq:mod}
|\braket{\psi_w}{\psi_i}|^2 = \frac{1}{2^n} + \frac{1}{2^{2n-1}}\sum_{i<j}^{2^n-1}\cos((\theta_j - \phi_j) - (\theta_i - \phi_i))\ .
\end{equation}
It is easily checked that   $|\braket{\psi_w}{\psi_i}|^2=1$ for $\theta_i=\phi_i\  \forall i$, since the two states would coincide in such case. 

Equation \eqref{eq:mod} represents the \textit{activation function} implemented by the proposed quantum neuron. Even if it does not remind any of the activation functions conventionally used in classical machine learning techniques, such as the Sigmoid or ReLu functions \cite{DeepLearning}, its nonlinearity suffices to accomplish classification tasks, as we will discuss in the following sections. 

\subsubsection*{Color invariance and noise resilience}

From Eq.~\eqref{eq:mod}, we define the activation function of the quantum artificial neuron as
%\begin{eqnarray}
%f(\bm{\theta},\bm{\phi}) &=&\frac{1}{2^n} +  %\frac{1}{2^{2n-1}} \times \nonumber \\ 
%&&  \times \sum_{i<j}^{2^n-1}\cos((\theta_j - \phi_j) %- (\theta_i - \phi_i))\ .
%\label{eq:mod2}
%\end{eqnarray}
\begin{equation}
f(\bm{\theta},\bm{\phi}) =
|\braket{\psi_w}{\psi_i}|^2 \ .
\label{eq:mod2}
\end{equation}
Keeping $\bm{\phi}$ fixed, suppose two different input vectors are passed to the quantum neuron: $\bm{\theta}$ and $\bm{\theta'}=\bm{\theta}+\bm{\Delta}$, with $\bm{\Delta}=(\Delta,\hdots, \Delta)$. Whatever the value of $\Delta$, it is easy to infer that both  input vectors will result in the same activation function. Hence, two input vectors only differing by a constant, albeit real valued, quantity will be equally classified by such model of quantum perceptron. Hence, in the context of image classification, we can state that the present algorithm has a built in color translational invariance. This should not come as a surprise, since the activation function actually depends of the \textit{differences} between phases. In fact, the artificial neuron tends to recognize as similar any dataset that displays the same overall differences, instead of perfectly coincident datasets.

Next, we assume that the input and weight vectors do coincide, but only up to some noise corrupting the input vector, such that: $\bm{\theta} = \bm{\phi} + \bm{\Delta}$, where $\bm{\Delta}=(\Delta_0, \Delta_1, \hdots, \Delta_{2^n-1})$ represents the small variations, now assumed to be different on each pixel. Substituting the above values in Eq.~\eqref{eq:mod2}, we obtain
\begin{equation}
f(\bm{\theta},\bm{\phi}) = f(\bm{\Delta}) = \frac{1}{2^n} + \frac{1}{2^{2n-1}}\sum_{i<j}^{2^n-1}\cos(\Delta_j-\Delta_i)\ .
\end{equation}
Assuming then the noise factors, $\Delta_i$, distributed according to a uniform distribution in the interval $[-a/2, a/2]$, 
the  activation function averaged over the probability distribution of $\Delta_i$ can be calculated as (see App.~\ref{app:Noise}):
\begin{equation}
\langle  f(\bm{\Delta})  \rangle = \frac{1}{2^n}+\frac{2^n-1}{2^{n-1}}\left(\frac{1-\cos(a)}{a^2}\right)\ .
\label{eq:noise}
\end{equation}
Since all the possible input data lie in the interval $[0,\pi/2]$, a reasonable noise would be of the order of some fraction of $\pi/2$, which implies $a<1$. Hence, in the case of small noise, Eq.~\eqref{eq:noise} can be recast as
\begin{equation}
\langle  f(\bm{\Delta})  \rangle = 1- \frac{2^n-1}{2^n}\frac{a^2}{12} + O\left(a^6\right)\  \quad \text{for } a\ll1 \ .
\label{eq:noiseapprox}
\end{equation}
Thus, the classification of the quantum neuron is only slightly perturbed by the presence of noise corrupting an input vector otherwise having a perfect activation. By similar calculations, it can be shown that this property also holds for any kind of input vector, not only those with perfect activation (see App.~\ref{app:Noise}). \\
After having outlined the main steps defining the quantum perceptron model for continuously valued input vectors, we now proceed to build a quantum circuit that allows implementing it on a qubit-based quantum computing hardware.

\subsection{Quantum circuit model of a continuously valued perceptron}

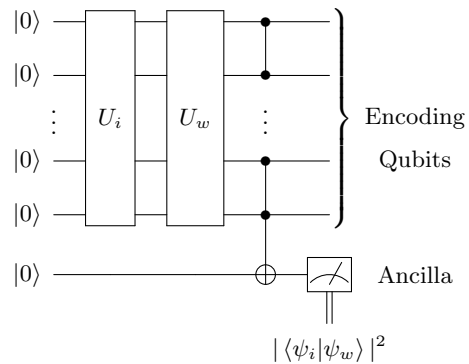
\begin{figure}[ht]
\[
\Qcircuit @C=1.3em @R=1.3em {
\lstick{\ket{0}} & \multigate{4}{{U_i}} & \multigate{4}{U_w} & \ctrl{1} & \qw \\ 
\lstick{\ket{0}} & \ghost{{U_i}} &  \ghost{{U_w}} & \ctrl{0} & \qw \\        
\raisebox{.3em}{\vdots} 	 &  &  & \raisebox{.3em}{\vdots} & & & \text{Encoding}	\\        
\lstick{\ket{0}} & \ghost{{U_i}} &  \ghost{{U_w}} & \ctrl{1} &  \qw & & \text{Qubits} \\        
\lstick{\ket{0}} & \ghost{{U_i}} &  \ghost{{U_w}} & \ctrl{1} & \qw 
\gategroup{1}{5}{5}{5}{1.0em}{\}}  \\
\lstick{\ket{0}} & 		\qw		  &  			\qw		   & \targ     & \meter & & \mbox{Ancilla} \\
& & & & \dstick{|\braket{\psi_i}{\psi_w}|^2} \cwx         
}
\]
\\
\caption{Quantum circuit model of a perceptron with continuously valued input and weight vectors.}
\label{fig:QN}
\end{figure}

A quantum circuit implementing the quantum neuron described above is schematically represented in Fig.~\ref{fig:QN}. The first section of the circuit, denoted as $U_i$, transforms the quantum register, initialized into the reference state $|0\rangle^{\otimes n}$, to the input quantum state defined in Eq.~\eqref{eq:input}; the following operation $U_w$ performs the inner product between the input and weight quantum state; finally, a multi-controlled CNOT targeting an ancillary qubit is used to extract the final result of the computation. We now explain in detail how these transformations can be achieved.

Starting from the $n$-qubit state, $\ket{00\hdots 0}=\ket{0}^{\otimes n}$, the $U_i$ operation creates the quantum input state $U_i\ket{0}^{\otimes n}=\ket{\psi_i}$ \eqref{eq:input}. Such a unitary can be built by means of a brute force approach. First of all, we apply a layer of Hadamard gates, H$^{\otimes n}$, which creates the balanced superposition state H$^{\otimes n}\ket{0}^{\otimes n}=\ket{+}^{\otimes n}$, with $\ket{+} = (\ket{0}+\ket{1})/\sqrt{2}$. The quantum state $\ket{+}^{\otimes n}$ consists of an equally weighted superposition of all the states in the $n$ qubits computational basis, hence we can target each of them and add the appropriate phase to it, in order to obtain the desired result. This action corresponds to the diagonal (in the computational basis) unitary operation
\begin{equation}
U(\bm{\theta}) := 
\begin{bmatrix}
e^{i\theta_0} & 0 & \cdots & 0 \\
0 & e^{i\theta_1}  & \cdots &  \\
\vdots  & \vdots  & \ddots & \vdots  \\
0 & 0 & \cdots & e^{i\theta_{2^n-1}} 
\end{bmatrix}
\label{eq:Ui_matrix}
\end{equation}
whose action is to phase shift each state of the computational basis, $\ket{i}$, to $e^{i\theta_i}\ket{i}$, with phases $\theta_i \in \mathbb{R}$, that are (in general) independent from each other. We decompose $$U(\bm{\theta})=\prod_{i=0}^{2^n-1} U(\theta_i)\ ,$$ where $U(\theta_i)$ is the unitary whose action is $U(\theta_i)\ket{i}=e^{i\theta_i}\ket{i}$, while leaving all the other states in the computational basis unchanged. These unitaries are equivalent to a combination of X gates and a multi-controlled phase shift gate, C$^{n-1}$R($\theta$), where the phase shift gate is the unitary operation defined as R($\theta$) = $\begin{bmatrix} 1 & 0 \\ 0 & e^{i\theta} \end{bmatrix}$ \cite{Nielsen_Chuang_2010}.

For example, suppose having $n=3$ qubits, and consider the state $\ket{101}$ to be phase shifted to $e^{i\theta_3}\ket{101}$. This transformation is achieved by the following quantum circuit:
\[
\Qcircuit @C=0.4em @R=0.8em {
& \qw & \ctrl{1}  & \qw &  &  & \qw & \ctrl{1} & \qw & \qw \\
& \qw & \ctrlo{1}  & \qw &
\push{\rule{.3em}{0em}=\rule{.3em}{0em}} & &
 \gate{X} & \ctrl{1} & \gate{X} & \qw \\
& \qw & \gate{\begin{bmatrix}1&0\\0&e^{i\theta_3}\end{bmatrix}} & \qw &  &  & \qw & \gate{\begin{bmatrix}1&0\\0&e^{i\theta_3}\end{bmatrix}} & \qw & \qw 
}
\]
%\[
%\Qcircuit @C=0.5em @R=1em {
%& \qw & \ctrl{1}  & \qw \\
%& \qw & \ctrlo{1}  & \qw \\
%& \qw & \gate{\begin{bmatrix}1&0\\0&e^{i\theta_5}\end{bmatrix}} & \qw & \qw 
%}
%\Qcircuit @C=0.5em @R=1em {
%& \qw & \ctrl{1} & \qw & \qw \\
%& \gate{X} & \ctrl{1} & \gate{X} & \qw \\
%& \qw & \gate{\begin{bmatrix}1&0\\0&e^{i\theta_5}\end{bmatrix}} & \qw & \qw 
%}
%\]
which implements the desired transformation $U(\theta_3)\ket{101}=e^{i\theta_3}\ket{101}$, while leaving all other states of the computational basis unchanged. Iterating a similar gate sequence for each state of the computational basis eventually yields the overall unitary operation, \eqref{eq:Ui_matrix}. So far, we have built the quantum circuit allowing to encode an arbitrary input vector: given the input $\vec{i}=(e^{i \theta_0}, e^{i \theta_2}, \cdots, e^{i\theta_{2^n-1}})$ as in Eq.~\eqref{eq:quantum_input}, we create the state $\ket{\psi_i}$ \eqref{eq:input}, by means of the operation $U_i\ket{0}^{\otimes n}=U(\bm{\theta})\text{H}^{\otimes n}\ket{0}^{\otimes n}=\ket{\psi_i}$, whose parameters depend on the input entries.

The unitary $U_w$ can then be constructed in a similar fashion. First of all, notice that the $U_i$ is unitary, thus reversible. Be $\vec{w}=(e^{i \phi_0}, e^{i \phi_2}, \cdots, e^{i \phi_{2^n-1}})$ the weight vector, then the desired inner product $\braket{\psi_w}{\psi_i}$ \eqref{eq:mod} resides in the overlap of the quantum state $\ket{\phi_{i,w}}=(U(\bm{\phi})\text{H}^{\otimes n})^\dagger \ket{\psi_i}$ with the ground state $\ket{0\hdots 0}$. In fact, since $U({\bm{\phi}})\text{H}^{\otimes n}\ket{0}^{\otimes n}=\ket{\psi_w}$ \eqref{eq:weight}, the scalar product is clearly given as
\begin{eqnarray}
\bra{0\hdots 0}\overbrace{(U(\bm{\phi})\text{H}^{\otimes n})^\dagger\ket{\psi_i}}^{\ket{\phi_{i,w}}} = \nonumber \\
\underbrace{\bra{0\hdots 0}\text{H}^{\otimes n}U(\bm{\phi})^\dagger}_{\bra{\psi_w}}\ket{\psi_i} = \nonumber \\
\braket{\psi_w}{\psi_i} \, . 
\label{eq:mid_step}
\end{eqnarray}

In order to extract the result, a final layer of X$^{\otimes n}$ gates is applied to all encoding qubits, such that the desired coefficient now multiplies the component $\ket{1}^{\otimes n}$ in the superposition:
\begin{eqnarray}
\text{X}^{\otimes n}\ket{\phi_{i,w}} &=& \ket{\bar{\phi}_{i,w}}   \nonumber \\
& = & \sum_{j=0}^{2^n-2} c_j \ket{j} + c_{2^n-1}\ket{11\hdots 1} 
\end{eqnarray}
with $c_{2^n-1}=\braket{\psi_w}{\psi_i}$. Thus, the $U_w$ transformation in Fig.~\ref{fig:QN} actually consists in the quantum operations $U_w=\text{X}^{\otimes n}\text{H}^{\otimes n}U(\bm{\phi})^\dagger$.

By means of a multi-controlled  C$^{n}$NOT, we load the result on an ancillary qubit
\begin{equation}
\text{C}^{n}\text{NOT}(\ket{\bar{\phi}_{i,w}}\ket{0}_a) = \sum_{j=0}^{2^n-1}c_j\ket{j}\ket{0}_a + c_{2^n-1} \ket{11\cdots 1}\ket{1}_a\ .
\end{equation}
Eventually, a final measurement of the ancilla qubit will yield result 1, which is interpreted as a firing neuron, with probability  $|c_{2^n-1}|^2=|\braket{\psi_w}{\psi_i}|^2=|\vec{i}\cdot\vec{w^*}|^2/(2^{2n})$, which consists in the neuron activation function, Eq.~\eqref{eq:mod}. \\

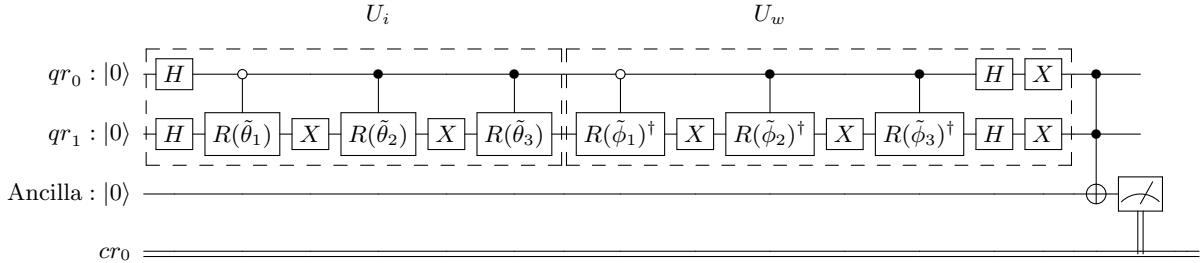
\begin{figure*}[ht]
\[
\Qcircuit @C=0.5em @R=0.7em @!R {
& & & & \mbox{$U_i$} & & & & & & \mbox{$U_w$} \\
\lstick{{qr}_{0} : \ket{0}} & \gate{H} & \ctrlo{1}  & \qw & \ctrl{1} & \qw & \ctrl{1} \push{\rule{1em}{1em}} & \qw & \ctrlo{1} & \qw & \ctrl{1} & \qw & \ctrl{1} & \gate{H} & \gate{X} & \qw & \ctrl{1} & \qw \\
\lstick{{qr}_{1} : \ket{0}} & \gate{H} & \gate{R(\tilde{\theta}_1)}  & \gate{X} & \gate{R(\tilde{\theta}_2)} & \gate{X} & \gate{R(\tilde{\theta}_3)} & \push{\rule{0em}{1.2em}} \qw \gategroup{2}{2}{3}{7}{0.8em}{--} & \gate{R(\tilde{\phi}_1)^\dagger} & \gate{X} & \gate{R(\tilde{\phi}_2)^\dagger} & \gate{X} & \gate{R(\tilde{\phi}_3)^\dagger} & \gate{H} & \gate{X} & \qw \gategroup{2}{9}{3}{15}{0.8em}{--} & \ctrl{1} & \qw \\
\lstick{{\text{Ancilla} : \ket{0}}} & \qw & \qw & \qw & \qw & \qw & \qw & \qw & \qw & \qw & \qw & \qw & \qw & \qw & \qw & \qw & \targ & \meter \cwx[1] \\
\lstick{cr_0} & \cw & \cw &  \cw & \cw & \cw &  \cw & \cw & \cw &  \cw & \cw & \cw &  \cw & \cw & \cw &  \cw & \cw & \cw &  \cw & \cw & \cw
}
\]
\caption{Scheme of the quantum circuit for the $n=2$ qubits case. The parameters are redefined as $\tilde{\theta}_i=\theta_i-\theta_0,\ \tilde{\phi_i}=\phi_i-\phi_0$, as detailed in Eq.~\eqref{eq:few_gates}.}
\label{fig:QCirc}
\end{figure*}

We notice that an input vector containing $N=2^n$ elements only requires $n+1$ qubits to implement the quantum circuit above, one of them being the ancilla qubit. To avoid introducing an ancilla qubit, an alternative strategy would be to perform a joint measurement on all $n$ qubits in the state $\ket{\phi_{i,w}}$ given in Eq.~\eqref{eq:mid_step}, with the probability of obtaining $\ket{0\hdots 0}$ being proportional to the inner product. However, with the idea of implementing the quantum computing version of a feedforward neural network, it is essential to have a model for which information is easily transferred from each neuron to the following layer. This can be accomplished by using an ancilla qubit per artificial neuron, where the quantity of interest can be loaded \cite{Tacchino1}. The time complexity of this quantum circuit depends linearly on the dimension of the input vectors $N$. Indeed, the quantum circuit introduced above requires $O(N)$ operations to implement all the phase shifts necessary to build the LME states, like Eq.~\eqref{eq:input}. Depending of the relation between the input data, $\theta_i$, other preparation schemes involving less operations could be devised \citep{Kraus}. 
Finally, it is worth noticing that thanks to global phase invariance, the activation function \eqref{eq:mod} can be recast as:
\begin{eqnarray} 
|\braket{\psi_w}{\psi_i}|^2 & = &\frac{1}{2^{2n}}\bigg| \displaystyle{\sum_{i=0}^{2^n-1}} e^{i(\theta_i-\phi_i)} \bigg|^2 \nonumber \\
& = & \frac{1}{2^{2n}}\bigg| 1 + \displaystyle{\sum_{i=1}^{2^n-1}} e^{i\left(\tilde{\theta}_i-\tilde{\phi}_i \right)} \bigg|^2\ , 
\label{eq:few_gates}
\end{eqnarray}
with $\tilde{\theta}_i=\theta_i-\theta_0,\ \tilde{\phi_i}=\phi_i-\phi_0$. By exploiting this redefinition of the parameters, it is possible to implement the same transformation but employing fewer gates, since it is equivalent to leaving the state $\ket{0}^{\otimes n}$ unchanged during the whole computation. Depending on the actual quantum hardware and data, further simplifications to the circuit could be obtained in compiling time. In Fig.~\ref{fig:QCirc}, the scheme of a quantum circuit implementing the artificial neuron model is shown for the specific case involving $n=2$ qubits.

%%%%%

\section{ Results: image recognition and learning }

\begin{figure}[H]
\centering
\begin{tikzpicture}[thick]
\begin{scope}[shift={(-7.25,1)},rotate=0,scale=1]
\draw [fill=white] (0,0) rectangle (2,2);
\draw [fill=white] (0,0) rectangle (1,1);
\draw [fill=white] (1,1) rectangle (2,2);
\node at (0.5,1.5)  {$255$};
\node at (1.5,1.5)  {$170$};
\node at (0.5,0.5)  {$85$};
\node at (1.5,0.5)  {$0$};
\end{scope}
\begin{scope}[shift={(-3.5,1)},rotate=0,scale=1]
\draw [fill=white] (0,0) rectangle (2,2);
\draw [fill=g170] (1,1) rectangle (2,2);
\draw [fill=g85] (0,0) rectangle (1,1);
\draw [fill=black] (1,0) rectangle (2,1);
\end{scope}
\end{tikzpicture}
\caption{A grayscale image with corresponding pixel intensities. This image can be encoded in the array $(255, 170, 85, 0)$, ordered from top-left to bottom-right.}
\label{fig:grayscale_img}
\end{figure}
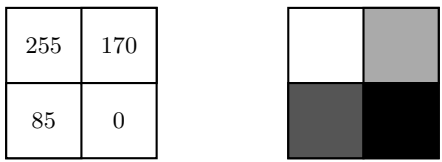

The quantum neuron model introduced above is an ideal candidate to perform classification tasks involving grayscale images. A grayscale image consists of a grid of pixels whose intensities are usually\footnote{This encoding of grayscale images employs a single byte (i.e., 8 bits) per pixel on a classical computing register.} represented by integer numbers in the range $[0, 255]$, as shown in Fig.~\ref{fig:grayscale_img}.

Since we make use of a phase encoding, all inputs (and weights) to the artificial neuron have to be normalized in the interval $[0, 2\pi]$. In this work we further restrict this domain for two reasons. First, values in $[0, \pi]$ and $[\pi, 2\pi]$ are fully equivalent, due to the periodicity in phase and the squared modulus in Eq.~\eqref{eq:mod}; second, for the same reason, states with zero or $\pi$ phase yield the same activation function, which in turn means that images with inverted colors (i.e., by exchanging white with black) would be recognized as equivalent by this perceptron model. Hence, to distinguish a given image from its negative, we further restrict the input and weight elements to lie in the range $[0, \pi/2]$.
Thus, an image such as the one reported in Fig.~\ref{fig:grayscale_img} is subject to the normalization $(255, 170, 85, 0) \rightarrow \frac{\pi/2}{255} (255, 170, 85, 0)$, before using it as an input vector to be encoded into the quantum neuron model. \\
We implemented and tested the quantum circuit both on simulators and on real quantum hardware, by using the IBM Quantum Experience~\footnote{https://quantum-computing.ibm.com/} and Qiskit~\cite{Qiskit}. The results are reported in the following.

\subsection{Numerical Results}

\begin{figure*}
\begin{tikzpicture}[thick]

\begin{scope}[shift={(-8.8,-0.6)},rotate=0,scale=1]
\draw [fill=white] (0,0) rectangle (2,2);
\draw [fill=black] (1,1) rectangle (2,2);
\draw [fill=black] (0,0) rectangle (1,1);
\draw [fill=white] (1,0) rectangle (2,1);
\node at (1.,2.3)  {$\bm{\phi}_{\text{target}}$};
\end{scope}

\begin{scope}[shift={(-6.6,1)},rotate=0,scale=1]
\draw [fill=white!91.76!black] (0,0) rectangle (2,2);
\draw [fill=white!14.11!black] (1,1) rectangle (2,2);
\draw [fill=white!12.15!black] (0,0) rectangle (1,1);
\draw [fill=white!33.72!black] (1,0) rectangle (2,1);
\node at (1,2.3)  {$9$};
\end{scope}

\begin{scope}[shift={(-4.4,1)},rotate=0,scale=1]
\draw [fill=g242] (0,0) rectangle (2,2);
\draw [fill=g60] (1,1) rectangle (2,2);
\draw [fill=g105] (0,0) rectangle (1,1);
\draw [fill=g242] (1,0) rectangle (2,1);
\node at (1.,2.3)  {$19$};
\end{scope}

%\begin{scope}[shift={(-4.4,1)},rotate=0,scale=1]
%\draw [fill=white!24.70!black] (0,0) rectangle (2,2);
%\draw [fill=white!16.47!black] (1,1) rectangle (2,2);
%\draw [fill=white!21.17!black] (0,0) rectangle (1,1);
%\draw [fill=white!56.47!black] (1,0) rectangle (2,1);
%\node at (1.,2.3)  {$14$};
%\end{scope}

%\begin{scope}[shift={(-2.2,1)},rotate=0,scale=1]
%\draw [fill=white!21.56!black] (0,0) rectangle (2,2);
%\draw [fill=white!34.50!black] (1,1) rectangle (2,2);
%\draw [fill=white!1.96!black] (0,0) rectangle (1,1);
%\draw [fill=white!98.43!black] (1,0) rectangle (2,1);
%\node at (1,2.3)  {$23$};
%\end{scope}

%\begin{tikzpicture}[thick]
%\begin{scope}[shift={(-8.8,1)},rotate=0,scale=1]
%\draw [fill=white!5.88!black] (0,0) rectangle (2,2);
%\draw [fill=white!25.49!black] (1,1) rectangle (2,2);
%\draw [fill=white!55.68!black] (0,0) rectangle (1,1);
%\draw [fill=white!20.78!black] (1,0) rectangle (2,1);
%\node at (1.,2.3)  {$3$};
%\end{scope}

\begin{scope}[shift={(-6.6,-1.8)},rotate=0,scale=1]
\draw [fill=white!6.66!black] (0,0) rectangle (2,2);
\draw [fill=white!90.19!black] (1,1) rectangle (2,2);
\draw [fill=white!76.86!black] (0,0) rectangle (1,1);
\draw [fill=white!43.13!black] (1,0) rectangle (2,1);
\node at (1,2.3)  {$7$};
\end{scope}

\begin{scope}[shift={(-4.4,-1.8)},rotate=0,scale=1]
\draw [fill=white!4.13!black] (0,0) rectangle (2,2);
\draw [fill=white!90.98!black] (1,1) rectangle (2,2);
\draw [fill=white!53.33!black] (0,0) rectangle (1,1);
\draw [fill=white!19.21!black] (1,0) rectangle (2,1);
\node at (1,2.3)  {$12$};
\end{scope}
%\begin{scope}[shift={(-2.2,1)},rotate=0,scale=1]
%\draw [fill=white!3.52!black] (0,0) rectangle (2,2);
%\draw [fill=white!78.82!black] (1,1) rectangle (2,2);
%\draw [fill=white!60.78!black] (0,0) rectangle (1,1);
%\draw [fill=white!72.54!black] (1,0) rectangle (2,1);
%\node at (1,2.3)  {$28$};
%\end{scope}

\node at (3.1,0.8) {\scalebox{0.3}{\input{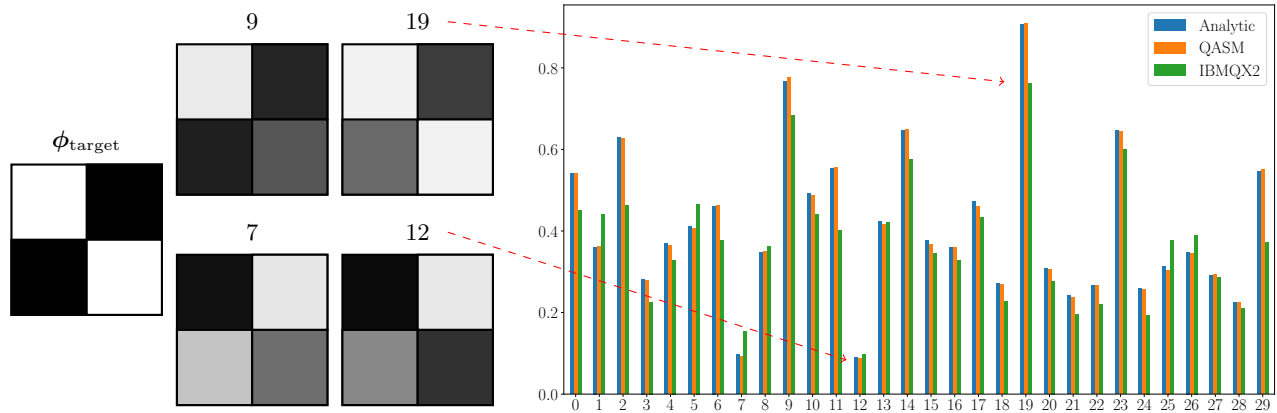}}};
\draw[dashed, thin,->, red] (-3.,3.3) -- (4.4,2.5);
\draw[dashed, thin,->, red] (-3.,0.5) -- (2.3,-1.2);

\end{tikzpicture}
\caption{Results of the image recognition task performed by a quantum artificial neuron, obtained by running the corresponding quantum circuit with either the  Qiskit QASM Simulator backend or the ibmqx2-Yorktown real quantum processor. In addition, we also report the related analytic values. The target weight vector, $\bm{\phi}_{\text{target}}$, is fixed, and $m=30$ random images are given as input vectors to the quantum artificial neuron. For each input, the corresponding quantum circuit is executed 8192 times. The checkerboard-like image corresponds to the target weight vector $\bm{\phi}_{\text{target}} = (\pi/2, 0, 0, \pi,2)$, while the images displaying respectively largest and lowest activation are the ones labeled as 19 and 12. Input vectors labeled as 9 and 7 are examples of images with high and low activation,  respectively.}
\label{fig:Comparison}
\end{figure*}

To better appreciate the potentialities of the continuously valued quantum neuron, we analyse its performance in recognizing similar images. We fix the weight vector to $\bm{\phi} = (\pi/2, 0, 0,  \pi/2)$, which corresponds to the checkerboard pattern represented in the image of Fig.~\ref{fig:Comparison}, and then generate a few random images to be used as inputs to the quantum neuron. For each input, the circuit is executed multiple times, thus building a statistics of the outcomes. With $m=30$ random generated images, the results of the classification are depicted in Fig.~\ref{fig:Comparison}, which includes the analytic results, the results of numerical simulations run on Qiskit QASM Simulator backend, and finally the results obtained by executing the quantum circuit on the ibmqx2-yorktown (accessed in March 2020) real device. Due to errors in the actual quantum processing device, the statistics of the outcome differ from either the simulated one or the analytic result. Nevertheless, the same overall behaviour can be easily recognised, thus showing that the quantum neuron circuit can be successfully implemented also in an actual quantum processor giving reliable results for such recognition tasks. The images producing the largest activation are the ones corresponding to input vectors similar to the checkerboard-like weight vector, which confirms the desired behaviour of the quantum neuron in recognizing similar images. On the contrary, the images with lowest activation are similar to the negative of the target weight vector, as desired. 

\section{Learning}

The process of finding the appropriate value for the weights to implement a given classification is called \textit{learning}, and it is generally based upon an optimization procedure in which a cost function is minimized by some gradient descent technique. Ideally, the minimum of the cost function corresponds to the targeted solution.

A simple learning task for our quantum neuron is to recognize a single given input. Starting from an input vector, $\bm{\theta}$, we aim at finding a  weight vector, $\bm{\phi}$, producing a high activation. Since the activation function for our quantum neuron is given in Eq.~\eqref{eq:mod}, we know that perfect activation can only be obtained when the input and weight vectors are exactly coincident,  $\bm{\theta}=\bm{\phi}$. This case can easily be checked numerically, by letting the neuron learn the right weights through a classical optimization technique.

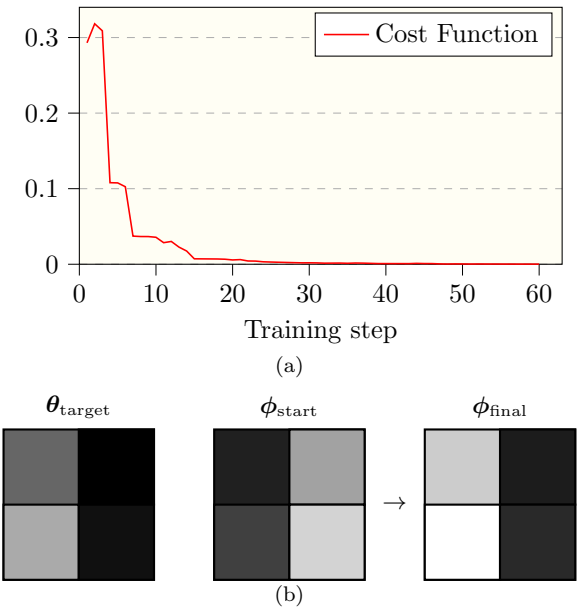
\begin{figure}

\captionsetup[subfloat]{farskip=2pt,captionskip=1pt}
\subfloat[\label{fig:learningSPSA}]{{\includegraphics{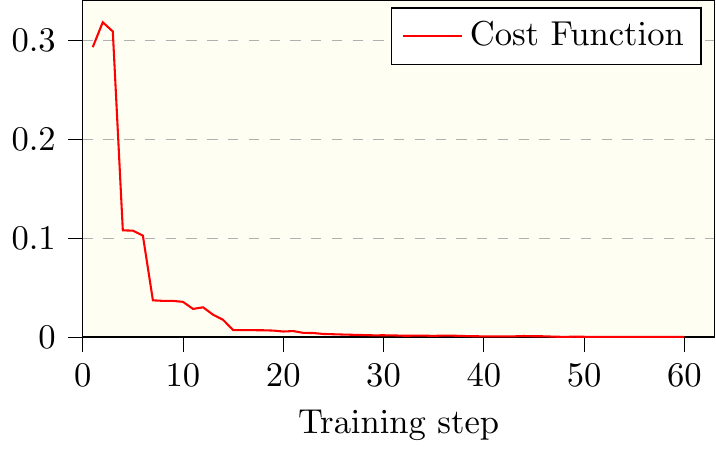}}}\hfill
\subfloat[\label{fig:finalW}]{
\begin{tikzpicture}[thick]
\begin{scope}[shift={(-8.8,1)},rotate=0,scale=1]
\draw [fill=g102] (0,0) rectangle (2,2);
\draw [fill=black] (1,1) rectangle (2,2);
\draw [fill=g170] (0,0) rectangle (1,1);
\draw [fill=g16] (1,0) rectangle (2,1);
\node at (1,2.3) {$\bm{\theta}_{\text{target}}$};
\end{scope}

\begin{scope}[shift={(-6,1)},rotate=0,scale=1]
\draw [fill=g32] (0,0) rectangle (2,2);
\draw [fill=g162] (1,1) rectangle (2,2);
\draw [fill=g65] (0,0) rectangle (1,1);
\draw [fill=g211] (1,0) rectangle (2,1);
\node at (1,2.3) {$\bm{\phi}_{\text{start}}$};
\node at (2.4,1) {$\rightarrow$};
\end{scope}

\begin{scope}[shift={(-3.2,1)},rotate=0,scale=1]
\draw [fill=g204] (0,0) rectangle (2,2);
\draw [fill=g28] (1,1) rectangle (2,2);
\draw [fill=white] (0,0) rectangle (1,1);
\draw [fill=g40] (1,0) rectangle (2,1);
\node at (1,2.3) {$\bm{\phi}_{\text{final}}$};
\end{scope}
\end{tikzpicture}
}
\caption{(a) Minimization of the cost function, $\mathcal{L}(\bm{\phi}) = (1-f(\bm{\theta}, \bm{\phi}))^2$, as a function of iteration steps. (b) Image corresponding to the target weight vector, $\bm{\theta}$ (left panel), and the weight vector before the optimization, $\bm{\phi}_0$ (center panel), and after the optimization, $\bm{\phi}_f$ (right panel).}
\label{fig:EasyLearning}
\end{figure}

A naive yet efficient choice for the cost function driving the learning process is $\mathcal{L}(\bm{\phi}) = (1-f(\bm{\theta}, \bm{\phi}))^2$, in which $f(\bm{\theta}, \bm{\phi})$ is the activation function of the artificial neuron with input $\vec{\theta}$ and weight vector $\bm{\phi}$, as in Eq.~\eqref{eq:mod2}. The minimum of the cost function, zero, is reached when the quantum neuron has full activation, i.e. $f(\bm{\theta}, \bm{\phi})=1$.
The minimization process is driven by the Stochastic Perturbation Stochastic Approximations (SPSA) \cite{SPSA}, which is built for optimization processes characterized by the presence of noise and is thus particularly effective in the presence of probabilistic measurement outputs. \\
An actual implementation on the QASM simulator leads to the following results. The task is to recognize the input vector $\vec{\theta}=(\pi/5,\ 0,\ \pi/3,\ 0.1)$. Using the SPSA optimizer, the cost function gets minimized by varying the weight vector, as reported in Fig.~\ref{fig:learningSPSA}, where it is evident that the cost function rapidly converges to values close to zero after a few iteration steps.
The solution to the problem, that is the final optimized weight vector, is $\vec{\phi}_f = (1.03,\ 0.19,\ 1.47,\ 0.61)$, whose grayscale representation is plotted in Figure \ref{fig:finalW}.
Even if the input and weight vector are not numerically equivalent, we can see that the final weight image actually looks very much like the target one, as expected. In fact, the two images retain almost the same shades of gray, with the optimized one being a bit shifted towards the brightest end of the spectrum, and as we previously noticed, the neuron is blind to overall color shifts.\\

In general, when dealing with a classification task, there is more than one input vector to be classified. Let us restrict ourselves to the case of a supervised binary classification\footnote{This can be generalized in the case of a multi-class classification, by adopting a \textit{one versus all} approach.}, where a each input $\bm{\theta}$ is associated to a binary label, $y$, such that $y \in \{0, 1\}$. Thus, the learning procedure consists in finding the right parameters (i.e. a weight vector $\vec{w}$) for which the artificial neuron reproduces the correct association of a given input vector with its corresponding label. In order to implement this dichotomy in the perceptron model, it is common practice to introduce a \textit{threshold} value, $t$: given an input and a weight vector, if the activation of the artificial neuron is above the value set by $t$, then the assigned label is $1$; otherwise it is $0$. 

A common choice for the cost function is the distance of the correct label assignment from the one implemented by the artificial neuron, which is expressed as
\begin{equation}
\mathcal{L}(\bm{\phi}) = \frac{1}{M}\sum_{i=1}^{M}(y_i - \tilde{y}_i)^2\ ,
\label{eq:cost_function}
\end{equation}
where $M$ is the number of input entries, $y_i$ is the correct label associated to input value $\bm{\theta}_i$, and $\tilde{y_i}$ is the label assigned by the neuron, which is calculated as
\begin{equation}
\tilde{y}_i = 
\begin{cases}
1\quad \text{if } f(\bm{\theta}, \bm{\phi})> t\ , \\
0\quad \text{otherwise}
\end{cases}\ .
\label{eq:assigned_label}
\end{equation}
The learning process then consists in minimizing the cost function, such as the one in Eq.~\eqref{eq:cost_function}.

Generally speaking, in a supervised learning procedure the inputs are divided into two distinct sets: the \textit{training} set, which contains the input values that are used to drive the learning procedure, and the \textit{test} set, which contains input vectors used to test the actual classification power of the quantum neuron with data never analysed before. Now that we have introduced the general learning framework, we can apply it to a few specific cases.

\subsection{Learning of two dimensional data}
\begin{figure*}
%\captionsetup[subfloat]{\subfigcapskip=-10pt}
\hspace{0em}
\subfloat[\label{fig:2Da}]{\scalebox{0.52}{\input{2Ddatas_original.pgf}}}\hspace{0em}%
\subfloat[\label{fig:2Db}]{\scalebox{0.52}{\input{2D_decisionb.pgf}}}\hspace{0em}%
\subfloat[\label{fig:2D_learning}]{{\includegraphics{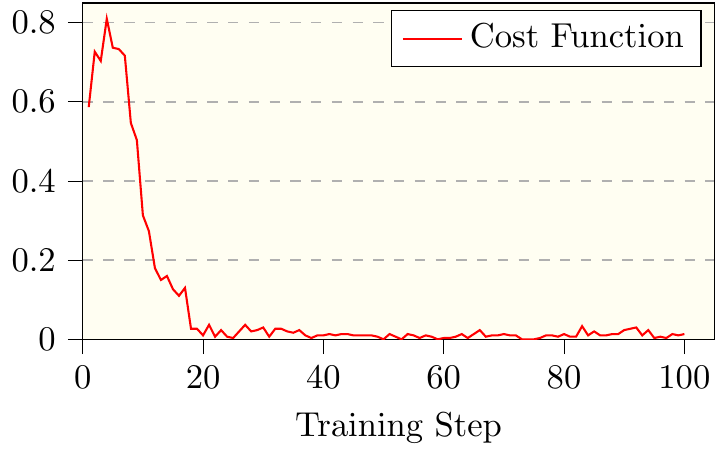}}}
\caption{Classification of two dimensional data. (a) Input data used as training set. (b) Test set and decision boundary implemented by the quantum neuron at the end of the learning procedure. The threshold used is $t=0.95$. (c) Optimization with the SPSA optimizer run on the QASM Simulator.}
\label{fig:2Ddata_original}
\end{figure*}

As a first example, we consider a classification problem of the form $\{\bm{x}_i, y_i\}_{i=1,\hdots, M}$, in which $\bm{x}_i=(x_1^i, x_2^i)$ are two dimensional input data, and $y_i$ their labels, such as the ones represented in Fig.~\ref{fig:2Da}. The color indicates the label associated to the input value, i.e., red for zero and blue for one.  
Since the data are two dimensional, we only need a single qubit to encode the information in the quantum state.
The cost function  \eqref{eq:cost_function} is minimized using the SPSA optimizer and its behaviour is reported in Fig.~\ref{fig:2D_learning}. The learning procedure converges towards a minimum of the cost function, and its value on the test set displayed in Fig.~\ref{fig:2Db} amounts to $\mathcal{L}_{\text{test}}=0$. This can be seen in Fig.~\ref{fig:2Db}, where we plot the decision boundary of the neuron along with the input values of the test set. All the calculations were performed on the QASM simulator. 

\subsection{Non separable points using a bias}

\begin{figure*}
\hspace{0em}
\subfloat[\label{fig:2DcircleA}]{\scalebox{0.52}{\input{2D_circle.pgf}}}\hspace{0em}
\subfloat[\label{fig:2DcircleB}]{\scalebox{0.52}{\input{2DCircle_decisionb.pgf}}}\hspace{0em}
\subfloat[\label{fig:Circ_learning}]{{\includegraphics{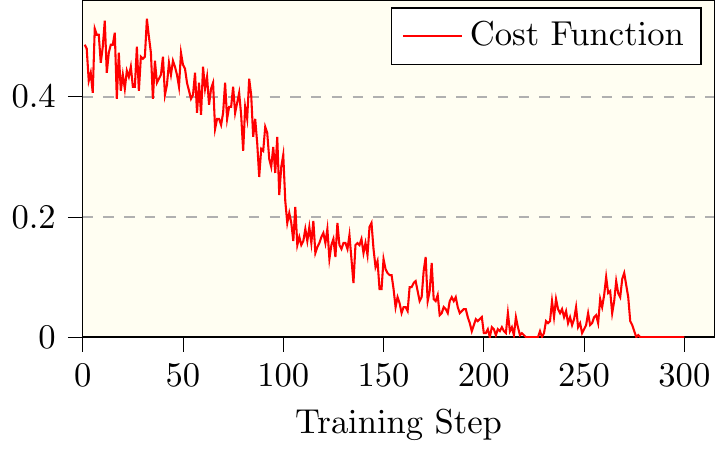}}}
\caption{Classification of two dimensional circles. (a) Input data used as the training set. (b) Test set and decision boundary implemented by the neuron at the end of the learning procedure. The threshold used in this example is $t=0.95$, and the bias $b=0.25$. (c) Optimization by the SPSA optimizer run on the QASM Simulator. The supervised learning procedure was performed with a batch examples of size 20, which explains why the error is not smooth but presents several spikes.}
\label{fig:Circ_data}
\end{figure*}

We have just shown that a single neuron is sufficient to classify some kind of two dimensional data. The procedure might fail on more complex structures of the dataset, though. For example, if one needs to classify data as in Fig.~\ref{fig:2DcircleA}, a single qubit encoding of the quantum perceptron model is not enough. However, using a quantum neuron implemented with two qubits allows to capture more degrees of freedom, thus helping to successfully tackle the problem. In fact, with $n=2$ qubits it is possible to encode $2^2=4$ parameters, or input data. Two of these are employed to encode the actual data of interest, one can be kept fixed to zero\footnote{Looking at the form of the activation function in \eqref{eq:mod}, it can be seen that it only depends on the \textit{differences} between the parameters. Thus, fixing one of the parameters to a constant value can be thought as just choosing a reference point.}, and the last free parameter can be interpreted as a \textit{bias}. Thus, a convenient encoding scheme is to consider input vectors of the form $\bm{\theta}=(\theta_0, \theta_1, \theta_2, \theta_3) = (0, x_1, x_2, 0)$, and weight vectors $\bm{\phi}=(\phi_0, \phi_1, \phi_2, \phi_3)=(0, \phi_1, \phi_2, b)$, where $b$ denotes the bias.
After the learning procedure, reported in Fig.~\ref{fig:Circ_learning}, the test set is classified as in Fig.~\ref{fig:2DcircleB}.

\subsection{MNIST dataset}

\begin{figure}[h]
\centering
%\subfloat[An exemple of a "one" from the MNIST dataset. \label{fig:MNIST0}]
{\includegraphics[width=0.225\textwidth]{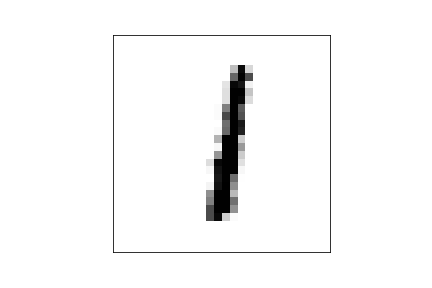}}%\hfill
%\subfloat[An exemple of a "zero" from the MNIST dataset.\label{fig:MNIST1}]
{\includegraphics[width=0.225\textwidth]{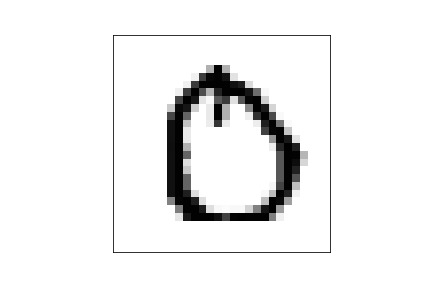}}%\hfill
\caption{Examples of images drawn from the MNIST dataset.}
\label{fig:MNIST}
\end{figure}

\begin{figure}[h]
\centering
\scalebox{0.7}{\input{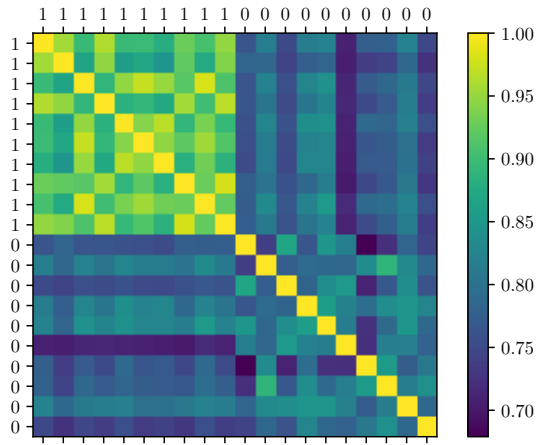}}
\caption{Confusion matrix related to some of the sample ``one'' and ``zero'' images taken from the MNIST dataset, evaluated with the activation function in Eq.~\eqref{eq:mod} and implemented by our quantum neuron model.}
\label{fig:MNIST_confusion}
\end{figure}

As a concluding example, it is interesting to show the application of the proposed quantum neuron model to the classification of the MNIST dataset, composed of $70000$ grayscale images of digits ranging from zero to nine. A selection of  sample images extracted from the given dataset are reported in Fig.~\ref{fig:MNIST}. We limit ourselves to the binary problem of correctly classifying the images of zeros and ones. Since each image in the MNIST dataset is composed of $28\times 28$ pixels, which is clearly not in the suited form $2^{n/2} \times 2^{n/2}$ required to be encoded on the quantum state of an artificial neuron with $N=2^n$ input data, we modify the images by adding a number of white redundant pixels, such that the processed images have $32\times 32$ pixels. A quantum artificial neuron with $n=10$ qubits can thus be used to encode the input images.
Here  we limit our analysis to checking whether the activation function introduced in Eq.~\eqref{eq:mod} is sufficient to discriminate between the encoded images of zeros and ones. With this goal in mind, we fix the weight vector of the artificial neuron to a sample ``one" selected  from the MNIST dataset, and then proceed to the classification with the remaining input images. Using a threshold of $t=0.85$ in Eq.~\eqref{eq:assigned_label}, the cost function evaluated on a test set of $m=2060$ images amounts to $\mathcal{L}\sim 0.02$, which in turn means an accuracy $\sim98\%$. In Fig.~\ref{fig:MNIST_confusion} it is shown the confusion matrix of some zeros and ones from the MNIST dataset evaluated with the activation function of the quantum neuron. According to the artificial neuron, the ``ones" are more similar among each other with respect to the ``zeros". Even if classical machine learning techniques can yield a classification accuracy above $99\%$, the present results show a remarkable degree of precision, also considering that in this particular example no learning and optimization procedure has been used, and  just a single quantum neuron has been used for the classification. In addition to this strategy, we also tried a \textit{pooling} procedure, in which each image in the MNIST dataset is first reduced to a $4\times 4$ image by means of a \textit{mean pooling} filter, and then classified by the neuron. After the learning, the neuron reaches a best accuracy around $80\%$. Nonetheless, these preliminary results show the potential of the activation function implemented by the quantum neuron to be used for recognition of complex patterns, such as numerical digits. Our quantum neuron model performs well when compared with other proposed quantum algorithms for the classification of the MNIST dataset. In fact, alternative algorithms have been proposed for this task, some of them using a hybrid classical-quantum approach, such as leveraging well established classical pre/post processing of data through classical machine learning techniques \cite{Qiskit-Textbook, MNIST_Kerenidis}. These hybrid approaches may yield higher (although comparable) classification accuracy when compared to our quantum neuron model. However, we emphasize that in our case the artificial neuron model is fully quantum in nature. When compared to other works using only quantum resources, our model seems to yield better results \cite{MNIST_tensorflow, ClassificationMNIST}.

\section{CONCLUSIONS}
{We have reported on a novel quantum algorithm allowing to implement a generalized perceptron model on a qubit-based quantum register. This quantum artificial neuron accepts and analyzes continuously valued input data. The proposed algorithm is translated into a quantum circuit model to be readily run on existing quantum hardware. It takes full advantage of the exponentially large Hilbert space available to encode input data on the phases of large superposition states, known as locally maximally entanglable (LME). These LME states can be constructed with a bottom-up approach, by imprinting each single phase separately. However, it should be stressed that alternative and possibly more efficient strategies could directly yield such states as ground states of suitable Hamiltonians, or as stationary states from dissipative processes \cite{Kraus}. The proposed continuously valued quantum neuron proves to be a good candidate for classification tasks of linearly non-separable two dimensional data, mostly related to pattern recognition tasks involving grayscale images. In this regard, thanks to the phase encoding, the neuron can leverage a built-in ``color translational" invariance, as well as significant noise resilience. In particular, the activation function implemented by the quantum neuron yields very high accuracy in the order of $~98\%$ when used to discriminate between images of zeros and ones from the MNIST dataset, thus indicating the ability to distinguish also complex patterns.
A further step would be to consider multiple layers of connected quantum neurons to build a continuous quantum feed-forward neural network. In addition, it would be interesting to study the application of phase encoding to other quantum machine learning techniques, such as quantum autoencoders. An important future direction would also be to design approximate methods to perform the weight unitary transformation in a way which scales more favorably with the number of encoding qubits: this could be achieved, for example, by training suitable variational or adaptive quantum circuits.
}

\begin{acknowledgments}
This research was partly supported by the Italian Ministry of Education, University and Research (MIUR) through the ``Dipartimenti di Eccellenza Program (2018-2022)'', Department of Physics, University of Pavia, the PRIN Project ``INPhoPOL'' and the Quantera project QuICHE. We acknowledge ENI S.p.A. for having partially contributed to this project through the Framework agreement with Universita` di Pavia. We acknowledge use of the IBM Quantum Experience for this work. The views expressed are those of the authors and do not reflect the official policy or position of IBM company or the IBM-Q team.
\end{acknowledgments}

\appendix

\section{Modulus square}
\label{app:A}
The squared modulus of the collection of complex numbers $\{z_i = r_i e^{i\gamma_i}\in \mathcal{C} | \ i = 1, \hdots, N\}$, is given as 

\begin{eqnarray}
\left|\sum_{i=1}^N z_i\right|^2 & = & \left(\sum_{i=1}^N z_i\right) \left(\sum_{j=1}^N z^*_j \right)  = \sum_{i,j=1}^N z_i  z^*_j\ , \nonumber \\
& = &  \sum_{i=j}^N |z_i|^2 + \sum_{i \neq j}^N r_i r_j e^{i(\gamma_i-\gamma_j)}\ , \nonumber \\
& = & \sum_{i=j}^N r_i^2 + 2 \sum_{i<j}^N r_i r_j \cos(\gamma_j - \gamma_i)\ ,
\end{eqnarray}
where in the last line the following relation has been applied
\begin{equation}
e^{i x}+e^{-ix} = 2 \cos(x)\ .
\end{equation}

Setting $r_i=1/N$ and $\gamma_i=\theta_i-\phi_i$,  respectively, we finally get:
\begin{equation}
 \left| \sum_{i=1}^N\frac{e^{i(\theta_i-\phi_i)}}{N}\right|^2  = \frac{1}{N} + \frac{2}{N^2}\sum_{i<j}^{N}\cos((\theta_j - \phi_j) - (\theta_i - \phi_i))\ ,
\end{equation}
which correctly reduces to Eq.~\eqref{eq:mod} in the main text, upon substituting $N=2^n$ and shifting the summation indices to start from zero. 

\section{Noise resilience}
\label{app:Noise}
Consider an input vector $\bm{\theta}$ equal to the weight vector $\bm{\phi}=\bm{\theta}$. Now, suppose that the input is corrupted and transformed into $\bm{\theta}'=\bm{\theta}+\bm{\Delta}$, with $\Delta = (\Delta_0, \Delta_1,\hdots, \Delta_{2^n-1})$. In this case, the activation function in Eq.~\eqref{eq:mod} reduces to
\begin{equation}
f(\bm{\Delta}) = \frac{1}{2^n} + \frac{1}{2^{2n-1}}\sum_{i<j}^{2^n-1}\cos(\Delta_j-\Delta_i)\ .
\end{equation}
Assuming that the noise values $\Delta_i$ are given by the uniform distribution:
\begin{equation}
\Delta_i \sim 
\begin{cases}
\frac{1}{a} \quad \text{for } x \in [-\frac{a}{2}, \frac{a}{2}] \\
0 \quad \text{otherwise}
\end{cases}
\quad \forall i\ , a \in \mathbb{R}\ ,
\label{eq:Uniform}
\end{equation}
it is possible to evaluate the \textit{average} activation function:
\begin{eqnarray}
\langle  f(\vec{\Delta})  \rangle & = & \frac{1}{2^n}+\frac{1}{2^{2n-1}}\sum_{i<j}^{2^n-1}\langle \cos(\Delta_j-\Delta_i) \rangle \nonumber \\
& = & \frac{1}{2^n}+\frac{1}{2^{2n-1}}\frac{2^n(2^n-1)}{2}\langle \cos(\Delta_j-\Delta_i) \rangle
\label{eq:App_noise0}
\end{eqnarray}
where in the last line it is implicitly assumed that $\langle \cos(\Delta_j-\Delta_i) \rangle$ is the same for all $i, j$. The averaging then yields
\begin{eqnarray}
\langle \cos(\Delta_j-\Delta_i) \rangle & = & \int_{-\frac{a}{2}}^{\frac{a}{2}} \int_{-\frac{a}{2}}^{\frac{a}{2}} \cos(\Delta_j-\Delta_i) \frac{d\Delta_i d \Delta_j}{a^2}\nonumber \\
& = & 2\left( \frac{1-\cos(a)}{a^2}\right)\ ,
\end{eqnarray}
and substituting back into \eqref{eq:App_noise0}, we  eventually get:
\begin{equation}
\langle  f(\vec{\Delta})  \rangle = \frac{1}{2^n}+\frac{2^n-1}{2^{n-1}}\left(\frac{1-\cos(a)}{a^2}\right)\ .
\end{equation}

Consider now the case where the input and weight vectors are different, $\bm{\theta}\neq \bm{\phi}$. The question is how much does the activation function change, if the input is corrupted by the presence of noise. As before, considering an input $\bm{\theta}'=\bm{\theta}+\bm{\Delta}$, the activation function reads:
\begin{equation}
f(\bm{\theta},\bm{\phi},\bm{\Delta}) = \frac{1}{2^n} + \frac{1}{2^{2n-1}}\sum_{i<j}^{2^n-1}\cos(A_{ij} + D_{ij})\ .
\end{equation}
with $A_{ij}=(\theta_j-\phi_j)-(\theta_i-\phi_i)$, and $D_{ij}=\Delta_j-\Delta_i$. Since $\cos(A_{ij}+D_{ij})=\cos(A_{ij})\cos(D_{ij})+\sin(A_{ij})\sin(D_{ij})$, and $\langle \sin(D_{ij})\rangle=0$ (using the probability distribution in Eq.~\eqref{eq:Uniform}), it finally results in
\begin{equation}
\langle f(\bm{\theta},\bm{\phi},\bm{\Delta})\rangle = \frac{1}{2^n} + \frac{D}{2^{2n-1}}\sum_{i<j}^{2^n-1}\cos(A_{ij})\ .
\end{equation}
with $D=2\left( \frac{1-\cos(a)}{a^2}\right)$.

% The \nocite command causes all entries in a bibliography to be printed out
% whether or not they are actually referenced in the text. This is appropriate
% for the sample file to show the different styles of references, but authors
% most likely will not want to use it.
%\nocite{*}

%\bibliographystyle{apsrev4}
\bibliography{cqn.bib}% Produces the bibliography via BibTeX.

%merlin.mbs apsrev4-1.bst 2010-07-25 4.21a (PWD, AO, DPC) hacked
%Control: key (0)
%Control: author (8) initials jnrlst
%Control: editor formatted (1) identically to author
%Control: production of article title (-1) disabled
%Control: page (0) single
%Control: year (1) truncated
%Control: production of eprint (0) enabled
\providecommand{\noopsort}[1]{}\providecommand{\singleletter}[1]{#1}%
\begin{thebibliography}{37}%
\makeatletter
\providecommand \@ifxundefined [1]{%
 \@ifx{#1\undefined}
}%
\providecommand \@ifnum [1]{%
 \ifnum #1\expandafter \@firstoftwo
 \else \expandafter \@secondoftwo
 \fi
}%
\providecommand \@ifx [1]{%
 \ifx #1\expandafter \@firstoftwo
 \else \expandafter \@secondoftwo
 \fi
}%
\providecommand \natexlab [1]{#1}%
\providecommand \enquote  [1]{``#1''}%
\providecommand \bibnamefont  [1]{#1}%
\providecommand \bibfnamefont [1]{#1}%
\providecommand \citenamefont [1]{#1}%
\providecommand \href@noop [0]{\@secondoftwo}%
\providecommand \href [0]{\begingroup \@sanitize@url \@href}%
\providecommand \@href[1]{\@@startlink{#1}\@@href}%
\providecommand \@@href[1]{\endgroup#1\@@endlink}%
\providecommand \@sanitize@url [0]{\catcode `\\12\catcode `\$12\catcode
  `\&12\catcode `\#12\catcode `\^12\catcode `\_12\catcode `\%12\relax}%
\providecommand \@@startlink[1]{}%
\providecommand \@@endlink[0]{}%
\providecommand \url  [0]{\begingroup\@sanitize@url \@url }%
\providecommand \@url [1]{\endgroup\@href {#1}{\urlprefix }}%
\providecommand \urlprefix  [0]{URL }%
\providecommand \Eprint [0]{\href }%
\providecommand \doibase [0]{http://dx.doi.org/}%
\providecommand \selectlanguage [0]{\@gobble}%
\providecommand \bibinfo  [0]{\@secondoftwo}%
\providecommand \bibfield  [0]{\@secondoftwo}%
\providecommand \translation [1]{[#1]}%
\providecommand \BibitemOpen [0]{}%
\providecommand \bibitemStop [0]{}%
\providecommand \bibitemNoStop [0]{.\EOS\space}%
\providecommand \EOS [0]{\spacefactor3000\relax}%
\providecommand \BibitemShut  [1]{\csname bibitem#1\endcsname}%
\let\auto@bib@innerbib\@empty
%</preamble>
\bibitem [{\citenamefont {Arute}\ \emph {et~al.}(2019)\citenamefont {Arute},
  \citenamefont {Arya}, \citenamefont {Babbush} \emph
  {et~al.}}]{QuantumSupremacy}%
  \BibitemOpen
  \bibfield  {author} {\bibinfo {author} {\bibfnamefont {F.}~\bibnamefont
  {Arute}}, \bibinfo {author} {\bibfnamefont {K.}~\bibnamefont {Arya}},
  \bibinfo {author} {\bibfnamefont {R.}~\bibnamefont {Babbush}},  \emph
  {et~al.},\ }\href {https://www.nature.com/articles/s41586-019-1666-5}
  {\bibfield  {journal} {\bibinfo  {journal} {Nature}\ }\textbf {\bibinfo
  {volume} {574}},\ \bibinfo {pages} {505–510} (\bibinfo {year}
  {2019})}\BibitemShut {NoStop}%
\bibitem [{\citenamefont {Preskill}(2018)}]{NISQ}%
  \BibitemOpen
  \bibfield  {author} {\bibinfo {author} {\bibfnamefont {J.}~\bibnamefont
  {Preskill}},\ }\href {\doibase 10.22331/q-2018-08-06-79} {\bibfield
  {journal} {\bibinfo  {journal} {{Quantum}}\ }\textbf {\bibinfo {volume}
  {2}},\ \bibinfo {pages} {79} (\bibinfo {year} {2018})}\BibitemShut {NoStop}%
\bibitem [{\citenamefont {Biamonte}\ \emph {et~al.}(2017)\citenamefont
  {Biamonte}, \citenamefont {Wittek}, \citenamefont {Pancotti}, \citenamefont
  {Rebentrost}, \citenamefont {Wiebe},\ and\ \citenamefont
  {Lloyd}}]{biamonte_quantum_2017}%
  \BibitemOpen
  \bibfield  {author} {\bibinfo {author} {\bibfnamefont {J.}~\bibnamefont
  {Biamonte}}, \bibinfo {author} {\bibfnamefont {P.}~\bibnamefont {Wittek}},
  \bibinfo {author} {\bibfnamefont {N.}~\bibnamefont {Pancotti}}, \bibinfo
  {author} {\bibfnamefont {P.}~\bibnamefont {Rebentrost}}, \bibinfo {author}
  {\bibfnamefont {N.}~\bibnamefont {Wiebe}}, \ and\ \bibinfo {author}
  {\bibfnamefont {S.}~\bibnamefont {Lloyd}},\ }\href@noop {} {\bibfield
  {journal} {\bibinfo  {journal} {Nature}\ }\textbf {\bibinfo {volume} {549}},\
  \bibinfo {pages} {195} (\bibinfo {year} {2017})}\BibitemShut {NoStop}%
\bibitem [{\citenamefont {Dunjko}\ and\ \citenamefont
  {Briegel}(2018)}]{Dunjko_Briegel_2018}%
  \BibitemOpen
  \bibfield  {author} {\bibinfo {author} {\bibfnamefont {V.}~\bibnamefont
  {Dunjko}}\ and\ \bibinfo {author} {\bibfnamefont {H.~J.}\ \bibnamefont
  {Briegel}},\ }\href {\doibase 10.1088/1361-6633/aab406} {\bibfield  {journal}
  {\bibinfo  {journal} {Reports on Progress in Physics}\ }\textbf {\bibinfo
  {volume} {81}},\ \bibinfo {pages} {074001} (\bibinfo {year}
  {2018})}\BibitemShut {NoStop}%
\bibitem [{\citenamefont {Benedetti}\ \emph {et~al.}(2019)\citenamefont
  {Benedetti}, \citenamefont {Lloyd}, \citenamefont {Sack},\ and\ \citenamefont
  {Fiorentini}}]{benedetti_parameterized_2019}%
  \BibitemOpen
  \bibfield  {author} {\bibinfo {author} {\bibfnamefont {M.}~\bibnamefont
  {Benedetti}}, \bibinfo {author} {\bibfnamefont {E.}~\bibnamefont {Lloyd}},
  \bibinfo {author} {\bibfnamefont {S.}~\bibnamefont {Sack}}, \ and\ \bibinfo
  {author} {\bibfnamefont {M.}~\bibnamefont {Fiorentini}},\ }\href@noop {}
  {\bibfield  {journal} {\bibinfo  {journal} {Quantum Science and Technology}\
  }\textbf {\bibinfo {volume} {4}},\ \bibinfo {pages} {043001} (\bibinfo {year}
  {2019})},\ \bibinfo {note} {publisher: IOP Publishing}\BibitemShut {NoStop}%
\bibitem [{\citenamefont {Schuld}\ \emph {et~al.}(2015)\citenamefont {Schuld},
  \citenamefont {Sinayskiy},\ and\ \citenamefont
  {Petruccione}}]{schuld_simulating_2015}%
  \BibitemOpen
  \bibfield  {author} {\bibinfo {author} {\bibfnamefont {M.}~\bibnamefont
  {Schuld}}, \bibinfo {author} {\bibfnamefont {I.}~\bibnamefont {Sinayskiy}}, \
  and\ \bibinfo {author} {\bibfnamefont {F.}~\bibnamefont {Petruccione}},\
  }\href@noop {} {\bibfield  {journal} {\bibinfo  {journal} {Physics Letters
  A}\ }\textbf {\bibinfo {volume} {379}},\ \bibinfo {pages} {660} (\bibinfo
  {year} {2015})}\BibitemShut {NoStop}%
\bibitem [{\citenamefont {Wiebe}\ \emph {et~al.}(2016)\citenamefont {Wiebe},
  \citenamefont {Kapoor},\ and\ \citenamefont {Svore}}]{wiebe_quantum_2016}%
  \BibitemOpen
  \bibfield  {author} {\bibinfo {author} {\bibfnamefont {N.}~\bibnamefont
  {Wiebe}}, \bibinfo {author} {\bibfnamefont {A.}~\bibnamefont {Kapoor}}, \
  and\ \bibinfo {author} {\bibfnamefont {K.~M.}\ \bibnamefont {Svore}},\
  }\href@noop {} {\bibfield  {journal} {\bibinfo  {journal} {arXiv:1602.04799
  [quant-ph, stat]}\ } (\bibinfo {year} {2016})}\BibitemShut {NoStop}%
\bibitem [{\citenamefont {Cao}\ \emph {et~al.}(2017)\citenamefont {Cao},
  \citenamefont {Guerreschi},\ and\ \citenamefont
  {Aspuru-Guzik}}]{cao_quantum_2017}%
  \BibitemOpen
  \bibfield  {author} {\bibinfo {author} {\bibfnamefont {Y.}~\bibnamefont
  {Cao}}, \bibinfo {author} {\bibfnamefont {G.~G.}\ \bibnamefont {Guerreschi}},
  \ and\ \bibinfo {author} {\bibfnamefont {A.}~\bibnamefont {Aspuru-Guzik}},\
  }\href@noop {} {\bibfield  {journal} {\bibinfo  {journal} {arXiv:1711.11240
  [quant-ph]}\ } (\bibinfo {year} {2017})}\BibitemShut {NoStop}%
\bibitem [{\citenamefont {Tacchino}\ \emph
  {et~al.}(2019{\natexlab{a}})\citenamefont {Tacchino}, \citenamefont
  {Macchiavello}, \citenamefont {Gerace},\ and\ \citenamefont
  {Bajoni}}]{Tacchino0}%
  \BibitemOpen
  \bibfield  {author} {\bibinfo {author} {\bibfnamefont {F.}~\bibnamefont
  {Tacchino}}, \bibinfo {author} {\bibfnamefont {C.}~\bibnamefont
  {Macchiavello}}, \bibinfo {author} {\bibfnamefont {D.}~\bibnamefont
  {Gerace}}, \ and\ \bibinfo {author} {\bibfnamefont {D.}~\bibnamefont
  {Bajoni}},\ }\href {\doibase 10.1038/s41534-019-0140-4} {\bibfield  {journal}
  {\bibinfo  {journal} {npj Quantum Information}\ }\textbf {\bibinfo {volume}
  {5}},\ \bibinfo {pages} {26} (\bibinfo {year}
  {2019}{\natexlab{a}})}\BibitemShut {NoStop}%
\bibitem [{\citenamefont {Torrontegui}\ and\ \citenamefont
  {Garcia-Ripoll}(2019)}]{torrontegui_unitary_2019}%
  \BibitemOpen
  \bibfield  {author} {\bibinfo {author} {\bibfnamefont {E.}~\bibnamefont
  {Torrontegui}}\ and\ \bibinfo {author} {\bibfnamefont {J.~J.}\ \bibnamefont
  {Garcia-Ripoll}},\ }\href@noop {} {\bibfield  {journal} {\bibinfo  {journal}
  {EPL (Europhysics Letters)}\ }\textbf {\bibinfo {volume} {125}},\ \bibinfo
  {pages} {30004} (\bibinfo {year} {2019})}\BibitemShut {NoStop}%
\bibitem [{\citenamefont {Kristensen}\ \emph {et~al.}(2019)\citenamefont
  {Kristensen}, \citenamefont {Degroote}, \citenamefont {Wittek}, \citenamefont
  {Aspuru-Guzik},\ and\ \citenamefont {Zinner}}]{kristensen_artificial_2019}%
  \BibitemOpen
  \bibfield  {author} {\bibinfo {author} {\bibfnamefont {L.~B.}\ \bibnamefont
  {Kristensen}}, \bibinfo {author} {\bibfnamefont {M.}~\bibnamefont
  {Degroote}}, \bibinfo {author} {\bibfnamefont {P.}~\bibnamefont {Wittek}},
  \bibinfo {author} {\bibfnamefont {A.}~\bibnamefont {Aspuru-Guzik}}, \ and\
  \bibinfo {author} {\bibfnamefont {N.~T.}\ \bibnamefont {Zinner}},\
  }\href@noop {} {\bibfield  {journal} {\bibinfo  {journal} {arXiv:1907.06269
  [cond-mat, physics:quant-ph]}\ } (\bibinfo {year} {2019})}\BibitemShut
  {NoStop}%
\bibitem [{\citenamefont {Havl{\'i}{\v c}ek}\ \emph {et~al.}(2019)\citenamefont
  {Havl{\'i}{\v c}ek}, \citenamefont {C{\'o}rcoles}, \citenamefont {Temme},
  \citenamefont {Harrow}, \citenamefont {Kandala}, \citenamefont {Chow},\ and\
  \citenamefont {Gambetta}}]{havlicek_supervised_2019}%
  \BibitemOpen
  \bibfield  {author} {\bibinfo {author} {\bibfnamefont {V.}~\bibnamefont
  {Havl{\'i}{\v c}ek}}, \bibinfo {author} {\bibfnamefont {A.~D.}\ \bibnamefont
  {C{\'o}rcoles}}, \bibinfo {author} {\bibfnamefont {K.}~\bibnamefont {Temme}},
  \bibinfo {author} {\bibfnamefont {A.~W.}\ \bibnamefont {Harrow}}, \bibinfo
  {author} {\bibfnamefont {A.}~\bibnamefont {Kandala}}, \bibinfo {author}
  {\bibfnamefont {J.~M.}\ \bibnamefont {Chow}}, \ and\ \bibinfo {author}
  {\bibfnamefont {J.~M.}\ \bibnamefont {Gambetta}},\ }\href@noop {} {\bibfield
  {journal} {\bibinfo  {journal} {Nature}\ }\textbf {\bibinfo {volume} {567}},\
  \bibinfo {pages} {209} (\bibinfo {year} {2019})}\BibitemShut {NoStop}%
\bibitem [{\citenamefont {Schuld}\ \emph {et~al.}(2017)\citenamefont {Schuld},
  \citenamefont {Fingerhuth},\ and\ \citenamefont
  {Petruccione}}]{schuld_implementing_2017}%
  \BibitemOpen
  \bibfield  {author} {\bibinfo {author} {\bibfnamefont {M.}~\bibnamefont
  {Schuld}}, \bibinfo {author} {\bibfnamefont {M.}~\bibnamefont {Fingerhuth}},
  \ and\ \bibinfo {author} {\bibfnamefont {F.}~\bibnamefont {Petruccione}},\
  }\href@noop {} {\bibfield  {journal} {\bibinfo  {journal} {EPL (Europhysics
  Letters)}\ }\textbf {\bibinfo {volume} {119}} (\bibinfo {year}
  {2017})}\BibitemShut {NoStop}%
\bibitem [{\citenamefont {Schuld}\ and\ \citenamefont
  {Killoran}(2019)}]{schuld_quantum_2019}%
  \BibitemOpen
  \bibfield  {author} {\bibinfo {author} {\bibfnamefont {M.}~\bibnamefont
  {Schuld}}\ and\ \bibinfo {author} {\bibfnamefont {N.}~\bibnamefont
  {Killoran}},\ }\href@noop {} {\bibfield  {journal} {\bibinfo  {journal}
  {Physical Review Letters}\ }\textbf {\bibinfo {volume} {122}},\ \bibinfo
  {pages} {040504} (\bibinfo {year} {2019})}\BibitemShut {NoStop}%
\bibitem [{\citenamefont {Romero}\ \emph {et~al.}(2017)\citenamefont {Romero},
  \citenamefont {Olson},\ and\ \citenamefont {Aspuru-Guzik}}]{Romero_2017}%
  \BibitemOpen
  \bibfield  {author} {\bibinfo {author} {\bibfnamefont {J.}~\bibnamefont
  {Romero}}, \bibinfo {author} {\bibfnamefont {J.~P.}\ \bibnamefont {Olson}}, \
  and\ \bibinfo {author} {\bibfnamefont {A.}~\bibnamefont {Aspuru-Guzik}},\
  }\href {\doibase 10.1088/2058-9565/aa8072} {\bibfield  {journal} {\bibinfo
  {journal} {Quantum Science and Technology}\ }\textbf {\bibinfo {volume}
  {2}},\ \bibinfo {pages} {045001} (\bibinfo {year} {2017})}\BibitemShut
  {NoStop}%
\bibitem [{\citenamefont {Lamata}\ \emph {et~al.}(2018)\citenamefont {Lamata},
  \citenamefont {Alvarez-Rodriguez}, \citenamefont {Mart{\'{\i}}n-Guerrero},
  \citenamefont {Sanz},\ and\ \citenamefont {Solano}}]{Lamata_2018}%
  \BibitemOpen
  \bibfield  {author} {\bibinfo {author} {\bibfnamefont {L.}~\bibnamefont
  {Lamata}}, \bibinfo {author} {\bibfnamefont {U.}~\bibnamefont
  {Alvarez-Rodriguez}}, \bibinfo {author} {\bibfnamefont {J.~D.}\ \bibnamefont
  {Mart{\'{\i}}n-Guerrero}}, \bibinfo {author} {\bibfnamefont {M.}~\bibnamefont
  {Sanz}}, \ and\ \bibinfo {author} {\bibfnamefont {E.}~\bibnamefont
  {Solano}},\ }\href {\doibase 10.1088/2058-9565/aae22b} {\bibfield  {journal}
  {\bibinfo  {journal} {Quantum Science and Technology}\ }\textbf {\bibinfo
  {volume} {4}},\ \bibinfo {pages} {014007} (\bibinfo {year}
  {2018})}\BibitemShut {NoStop}%
\bibitem [{\citenamefont {Henderson}\ \emph {et~al.}(2020)\citenamefont
  {Henderson}, \citenamefont {Shakya}, \citenamefont {Pradhan},\ and\
  \citenamefont {Cook}}]{henderson_quanvolutional_2020}%
  \BibitemOpen
  \bibfield  {author} {\bibinfo {author} {\bibfnamefont {M.}~\bibnamefont
  {Henderson}}, \bibinfo {author} {\bibfnamefont {S.}~\bibnamefont {Shakya}},
  \bibinfo {author} {\bibfnamefont {S.}~\bibnamefont {Pradhan}}, \ and\
  \bibinfo {author} {\bibfnamefont {T.}~\bibnamefont {Cook}},\ }\href@noop {}
  {\bibfield  {journal} {\bibinfo  {journal} {Quantum Machine Intelligence}\
  }\textbf {\bibinfo {volume} {2}},\ \bibinfo {pages} {1} (\bibinfo {year}
  {2020})}\BibitemShut {NoStop}%
\bibitem [{\citenamefont {Cong}\ \emph {et~al.}(2019)\citenamefont {Cong},
  \citenamefont {Choi},\ and\ \citenamefont {Lukin}}]{cong_quantum_2019}%
  \BibitemOpen
  \bibfield  {author} {\bibinfo {author} {\bibfnamefont {I.}~\bibnamefont
  {Cong}}, \bibinfo {author} {\bibfnamefont {S.}~\bibnamefont {Choi}}, \ and\
  \bibinfo {author} {\bibfnamefont {M.~D.}\ \bibnamefont {Lukin}},\ }\href@noop
  {} {\bibfield  {journal} {\bibinfo  {journal} {Nature Physics}\ }\textbf
  {\bibinfo {volume} {15}},\ \bibinfo {pages} {1273} (\bibinfo {year}
  {2019})}\BibitemShut {NoStop}%
\bibitem [{\citenamefont {Beach}\ \emph {et~al.}(2003)\citenamefont {Beach},
  \citenamefont {Lomont},\ and\ \citenamefont
  {Cohen}}]{Cohen_Quantum_Image_Processing_IEEE_2003}%
  \BibitemOpen
  \bibfield  {author} {\bibinfo {author} {\bibfnamefont {G.}~\bibnamefont
  {Beach}}, \bibinfo {author} {\bibfnamefont {C.}~\bibnamefont {Lomont}}, \
  and\ \bibinfo {author} {\bibfnamefont {C.}~\bibnamefont {Cohen}},\ }in\
  \href@noop {} {\emph {\bibinfo {booktitle} {Proceedings of 32nd Applied
  Imagery Pattern Recognition Workshop}}}\ (\bibinfo {year} {2003})\ pp.\
  \bibinfo {pages} {39--44}\BibitemShut {NoStop}%
\bibitem [{\citenamefont {Schuld}\ \emph {et~al.}(2014)\citenamefont {Schuld},
  \citenamefont {Sinayskiy},\ and\ \citenamefont
  {Petruccione}}]{Schuld_Petruccione_review_2014}%
  \BibitemOpen
  \bibfield  {author} {\bibinfo {author} {\bibfnamefont {M.}~\bibnamefont
  {Schuld}}, \bibinfo {author} {\bibfnamefont {I.}~\bibnamefont {Sinayskiy}}, \
  and\ \bibinfo {author} {\bibfnamefont {F.}~\bibnamefont {Petruccione}},\
  }\href {\doibase 10.1007/s11128-014-0809-8} {\bibfield  {journal} {\bibinfo
  {journal} {Quantum Information Processing}\ }\textbf {\bibinfo {volume}
  {13}},\ \bibinfo {pages} {2567} (\bibinfo {year} {2014})}\BibitemShut
  {NoStop}%
\bibitem [{\citenamefont {Zurada}(1992)}]{Zurada:intro_ANN_1992}%
  \BibitemOpen
  \bibfield  {author} {\bibinfo {author} {\bibfnamefont {J.~M.}\ \bibnamefont
  {Zurada}},\ }\href@noop {} {\emph {\bibinfo {title} {Introduction to
  Artificial Neural Systems}}}\ (\bibinfo  {publisher} {West Group},\ \bibinfo
  {year} {1992})\BibitemShut {NoStop}%
\bibitem [{\citenamefont {Rojas}(1996)}]{Rojas_ANN_Introduction}%
  \BibitemOpen
  \bibfield  {author} {\bibinfo {author} {\bibfnamefont {R.}~\bibnamefont
  {Rojas}},\ }\href@noop {} {\emph {\bibinfo {title} {Neural Networks: A
  Systematic Introduction}}}\ (\bibinfo  {publisher} {Springer},\ \bibinfo
  {year} {1996})\BibitemShut {NoStop}%
\bibitem [{\citenamefont {McCulloch}\ and\ \citenamefont
  {Pitts}(1943)}]{McCulloch_Pitts_1943}%
  \BibitemOpen
  \bibfield  {author} {\bibinfo {author} {\bibfnamefont {W.~S.}\ \bibnamefont
  {McCulloch}}\ and\ \bibinfo {author} {\bibfnamefont {W.}~\bibnamefont
  {Pitts}},\ }\href@noop {} {\bibfield  {journal} {\bibinfo  {journal} {The
  bulletin of mathematical biophysics}\ }\textbf {\bibinfo {volume} {5}},\
  \bibinfo {pages} {115} (\bibinfo {year} {1943})}\BibitemShut {NoStop}%
\bibitem [{\citenamefont {Rosenblatt}(1957)}]{Rosenblatt}%
  \BibitemOpen
  \bibfield  {author} {\bibinfo {author} {\bibfnamefont {F.}~\bibnamefont
  {Rosenblatt}},\ }\href@noop {} {\emph {\bibinfo {title} {The perceptron, a
  perceiving and recognizing automaton Project Para}}}\ (\bibinfo  {publisher}
  {Cornell Aeronautical Laboratory},\ \bibinfo {year} {1957})\BibitemShut
  {NoStop}%
\bibitem [{\citenamefont {Li}\ and\ \citenamefont {Xiao}(2013)}]{Li2013}%
  \BibitemOpen
  \bibfield  {author} {\bibinfo {author} {\bibfnamefont {P.}~\bibnamefont
  {Li}}\ and\ \bibinfo {author} {\bibfnamefont {H.}~\bibnamefont {Xiao}},\
  }\href {\doibase 10.1007/s10489-013-0447-3} {\bibfield  {journal} {\bibinfo
  {journal} {Applied Intelligence}\ }\textbf {\bibinfo {volume} {40}},\
  \bibinfo {pages} {107} (\bibinfo {year} {2013})}\BibitemShut {NoStop}%
\bibitem [{\citenamefont {Le}\ \emph {et~al.}(2010)\citenamefont {Le},
  \citenamefont {Dong},\ and\ \citenamefont
  {Hirota}}]{Hirota_quantum_images_polynomial_preparation}%
  \BibitemOpen
  \bibfield  {author} {\bibinfo {author} {\bibfnamefont {P.~Q.}\ \bibnamefont
  {Le}}, \bibinfo {author} {\bibfnamefont {F.}~\bibnamefont {Dong}}, \ and\
  \bibinfo {author} {\bibfnamefont {K.}~\bibnamefont {Hirota}},\ }\href
  {\doibase 10.1007/s11128-010-0177-y} {\bibfield  {journal} {\bibinfo
  {journal} {Quantum Information Processing}\ }\textbf {\bibinfo {volume}
  {10}},\ \bibinfo {pages} {63} (\bibinfo {year} {2010})}\BibitemShut {NoStop}%
\bibitem [{\citenamefont
  {Latorre}(2005)}]{Latorre_image_compression_entanglement}%
  \BibitemOpen
  \bibfield  {author} {\bibinfo {author} {\bibfnamefont {J.~I.}\ \bibnamefont
  {Latorre}},\ }\href@noop {} {\enquote {\bibinfo {title} {Image compression
  and entanglement},}\ } (\bibinfo {year} {2005}),\ \Eprint
  {http://arxiv.org/abs/arXiv:quant-ph/0510031} {arXiv:quant-ph/0510031}
  \BibitemShut {NoStop}%
\bibitem [{\citenamefont {Farhi}\ and\ \citenamefont
  {Neven}(2018)}]{ClassificationMNIST}%
  \BibitemOpen
  \bibfield  {author} {\bibinfo {author} {\bibfnamefont {E.}~\bibnamefont
  {Farhi}}\ and\ \bibinfo {author} {\bibfnamefont {H.}~\bibnamefont {Neven}},\
  }\href@noop {} {\enquote {\bibinfo {title} {Classification with quantum
  neural networks on near term processors},}\ } (\bibinfo {year} {2018}),\
  \Eprint {http://arxiv.org/abs/https://arxiv.org/abs/1802.06002}
  {https://arxiv.org/abs/1802.06002} \BibitemShut {NoStop}%
\bibitem [{\citenamefont {Kruszynska}\ and\ \citenamefont
  {Kraus}(2009)}]{Kraus}%
  \BibitemOpen
  \bibfield  {author} {\bibinfo {author} {\bibfnamefont {C.}~\bibnamefont
  {Kruszynska}}\ and\ \bibinfo {author} {\bibfnamefont {B.}~\bibnamefont
  {Kraus}},\ }\href {\doibase 10.1103/PhysRevA.79.052304} {\bibfield  {journal}
  {\bibinfo  {journal} {Phys. Rev. A}\ }\textbf {\bibinfo {volume} {79}},\
  \bibinfo {pages} {052304} (\bibinfo {year} {2009})}\BibitemShut {NoStop}%
\bibitem [{\citenamefont {LeCun}\ \emph {et~al.}(2015)\citenamefont {LeCun},
  \citenamefont {Bengio},\ and\ \citenamefont {Hinton}}]{DeepLearning}%
  \BibitemOpen
  \bibfield  {author} {\bibinfo {author} {\bibfnamefont {Y.}~\bibnamefont
  {LeCun}}, \bibinfo {author} {\bibfnamefont {Y.}~\bibnamefont {Bengio}}, \
  and\ \bibinfo {author} {\bibfnamefont {G.}~\bibnamefont {Hinton}},\ }\href
  {\doibase 10.1038/nature14539} {\bibfield  {journal} {\bibinfo  {journal}
  {Nature}\ }\textbf {\bibinfo {volume} {521}},\ \bibinfo {pages} {436–444}
  (\bibinfo {year} {2015})}\BibitemShut {NoStop}%
\bibitem [{\citenamefont {Nielsen}\ and\ \citenamefont
  {Chuang}(2010)}]{Nielsen_Chuang_2010}%
  \BibitemOpen
  \bibfield  {author} {\bibinfo {author} {\bibfnamefont {M.~A.}\ \bibnamefont
  {Nielsen}}\ and\ \bibinfo {author} {\bibfnamefont {I.~L.}\ \bibnamefont
  {Chuang}},\ }\href@noop {} {\emph {\bibinfo {title} {Quantum computation and
  quantum information}}},\ \bibinfo {edition} {10th}\ ed.\ (\bibinfo
  {publisher} {Cambridge University Press},\ \bibinfo {year}
  {2010})\BibitemShut {NoStop}%
\bibitem [{\citenamefont {Tacchino}\ \emph
  {et~al.}(2019{\natexlab{b}})\citenamefont {Tacchino}, \citenamefont
  {Barkoutsos}, \citenamefont {Macchiavello}, \citenamefont {Tavernelli},
  \citenamefont {Gerace},\ and\ \citenamefont {Bajoni}}]{Tacchino1}%
  \BibitemOpen
  \bibfield  {author} {\bibinfo {author} {\bibfnamefont {F.}~\bibnamefont
  {Tacchino}}, \bibinfo {author} {\bibfnamefont {P.}~\bibnamefont
  {Barkoutsos}}, \bibinfo {author} {\bibfnamefont {C.}~\bibnamefont
  {Macchiavello}}, \bibinfo {author} {\bibfnamefont {I.}~\bibnamefont
  {Tavernelli}}, \bibinfo {author} {\bibfnamefont {D.}~\bibnamefont {Gerace}},
  \ and\ \bibinfo {author} {\bibfnamefont {D.}~\bibnamefont {Bajoni}},\ }\href
  {https://arxiv.org/abs/1912.12486} {\bibfield  {journal} {\bibinfo  {journal}
  {arxiv:1912.12486}\ } (\bibinfo {year} {2019}{\natexlab{b}})}\BibitemShut
  {NoStop}%
\bibitem [{\citenamefont {Abraham}\ \emph {et~al.}(2019)\citenamefont {Abraham}
  \emph {et~al.}}]{Qiskit}%
  \BibitemOpen
  \bibfield  {author} {\bibinfo {author} {\bibfnamefont {H.}~\bibnamefont
  {Abraham}} \emph {et~al.},\ }\href {\doibase 10.5281/zenodo.2562110}
  {\enquote {\bibinfo {title} {Qiskit: An open-source framework for quantum
  computing},}\ } (\bibinfo {year} {2019})\BibitemShut {NoStop}%
\bibitem [{\citenamefont {{Spall}}(1998)}]{SPSA}%
  \BibitemOpen
  \bibfield  {author} {\bibinfo {author} {\bibfnamefont {J.~C.}\ \bibnamefont
  {{Spall}}},\ }\href@noop {} {\bibfield  {journal} {\bibinfo  {journal} {IEEE
  Transactions on Aerospace and Electronic Systems}\ }\textbf {\bibinfo
  {volume} {34}},\ \bibinfo {pages} {817} (\bibinfo {year} {1998})}\BibitemShut
  {NoStop}%
\bibitem [{\citenamefont {Asfaw}\ \emph {et~al.}(2020)\citenamefont {Asfaw},
  \citenamefont {Bello} \emph {et~al.}}]{Qiskit-Textbook}%
  \BibitemOpen
  \bibfield  {author} {\bibinfo {author} {\bibfnamefont {A.}~\bibnamefont
  {Asfaw}}, \bibinfo {author} {\bibfnamefont {L.}~\bibnamefont {Bello}},  \emph
  {et~al.},\ }\href {http://community.qiskit.org/textbook} {\enquote {\bibinfo
  {title} {Learn quantum computation using qiskit},}\ } (\bibinfo {year}
  {2020}),\ \bibinfo {note} {\uppercase{H}ybrid quantum-classical Neural
  Networks with PyTorch and Qiskit}\BibitemShut {NoStop}%
\bibitem [{\citenamefont {Kerenidis}\ and\ \citenamefont
  {Luongo}(2018)}]{MNIST_Kerenidis}%
  \BibitemOpen
  \bibfield  {author} {\bibinfo {author} {\bibfnamefont {I.}~\bibnamefont
  {Kerenidis}}\ and\ \bibinfo {author} {\bibfnamefont {A.}~\bibnamefont
  {Luongo}},\ }\href@noop {} {\enquote {\bibinfo {title} {Classification of
  mnist dataset with quantum slow feature analysis},}\ } (\bibinfo {year}
  {2018}),\ \Eprint {http://arxiv.org/abs/https://arxiv.org/abs/1805.08837}
  {https://arxiv.org/abs/1805.08837} \BibitemShut {NoStop}%
\bibitem [{\citenamefont {Broughton}\ \emph {et~al.}(2020)\citenamefont
  {Broughton}, \citenamefont {Verdon} \emph {et~al.}}]{MNIST_tensorflow}%
  \BibitemOpen
  \bibfield  {author} {\bibinfo {author} {\bibfnamefont {M.}~\bibnamefont
  {Broughton}}, \bibinfo {author} {\bibfnamefont {G.}~\bibnamefont {Verdon}},
  \emph {et~al.},\ }\href@noop {} {\bibfield  {journal} {\bibinfo  {journal}
  {arXiv:2003.02989 [cond-mat, physics:quant-ph]}\ } (\bibinfo {year}
  {2020})},\ \bibinfo {note} {\uppercase{M}NIST Classification
  Tutorial}\BibitemShut {NoStop}%
\end{thebibliography}%

\end{document}